\documentclass[10pt,a4paper]{article}

\usepackage{graphicx}
\usepackage[linesnumbered,tworuled,vlined]{algorithm2e}
\usepackage{amsmath}
\usepackage{amssymb,amsbsy}

\usepackage{pgf,pgfarrows,pgfnodes,tikz,color,tkz-berge}
\usetikzlibrary{arrows,shapes,snakes,automata,backgrounds,petri,topaths,calc}

\DeclareMathOperator*{\argmin}{\arg\!\min}

\newcommand{\vect}[1]{ \boldsymbol{  #1  } }

\newcommand{\R}{{\mathbb R}}
\newcommand{\Z}{{\mathbb Z}}

\newcommand{\D}[2]{ \ensuremath{ \frac{\mathrm{d} #1 }{\mathrm{d} #2 } }}

\newtheorem{definition}{Definition}[section]

\begin{document}

\title{A geometric method for model reduction of biochemical networks
with polynomial rate functions}

\author{Satya Swarup Samal$^1$, Dima Grigoriev$^2$,  Holger Fr\"ohlich$^1$,
Andreas Weber$^3$
\\ and Ovidiu Radulescu$^4$   \\
\small  $^1$ Algorithmic Bioinformatics, Bonn-Aachen International Center for IT,  Bonn, Germany,  \\
\small  $^2$ CNRS, Math\'ematiques, Universit\'e de Lille, Villeneuve d'Ascq,
 59655, France, \\
 \small $^3$ Institut f{\"u}r Informatik II, University of Bonn, Friedrich-Ebert-Allee 144, 53113 Bonn, Germany, \\
\small  $^4$ DIMNP UMR CNRS 5235, University of Montpellier, Montpellier, France.
 }

\maketitle

\tikzstyle{every picture}+=[remember picture]
\tikzstyle{na} = [baseline=-.5ex]

\centerline{\bf Abstract}

Model reduction of biochemical networks
relies on the knowledge of slow and fast variables.
We provide a geometric method, based on the Newton {polytope},
to identify slow variables of a biochemical network with polynomial rate functions.
The gist of the method is the notion of
tropical equilibration that provides
approximate descriptions
of slow invariant manifolds. Compared to extant
numerical algorithms such as the intrinsic low dimensional
manifold method, our approach is symbolic and
utilizes orders of magnitude instead of
precise values of the model parameters.
Application of this method to a large collection
of biochemical network models supports the idea that the number of dynamical variables in
minimal models of cell physiology can be small, in spite of the large number
of molecular regulatory actors.

\section{Introduction}
Model reduction is an important problem in computational biology. There are several methods
for reducing networks of biochemical reactions. Formal model reduction can be based on
 conservation laws, exact lumping \cite{Feret21042009}, and more generally, symmetry \cite{clarke1999model,rowley2000reconstruction}.
Approximate numerical reduction methods, such as computational singular perturbation
(CSP, \cite{lam1994csp}), intrinsic low dimensional  manifold (ILDM, \cite{maas1992simplifying})
 exploit the separation of timescales of various processes and variables.
 In dissipative systems, fast variables relax  rapidly to some low dimensional attractive
 manifold called invariant manifold \cite{gorban2005invariant} that carries the slow mode  dynamics.
 A projection of dynamical equations
 onto this manifold provides the reduced dynamics \cite{maas1992simplifying,gorban2005invariant}.
 This simple picture can be complexified to cope with hierarchies of invariant manifolds and
 with phenomena such as transverse instability, excitability and itineracy.
Firstly, the relaxation towards an attractor can have several stages, each with its own invariant manifold.
 During relaxation towards the attractor, invariant manifolds are usually embedded one into another (there is a decrease of dimensionality) \cite{chiavazzo2011adaptive}.
Secondly, invariant manifolds can lose local stability, which allow the trajectories to perform
large phase space excursions before returning in a different place on the same invariant manifold or on
a different one \cite{haller2010localized}. The set of slow variables can change from one place to another.
For all
 these reasons, even for fixed
 parameters, nonlinear models can have several reductions.

  CSP and ILDM methods provide numerical approximations of the invariant manifold close to an
 attractor. These methods have been successfully applied to reduce
 networks of reactions in chemical engineering.
 Other reduction methods utilize quasi-steady state {(QSS)} or
 quasi-equilibrium {(QE)} approximations \cite{GRZ10ces,radulescu2008robust,radulescu2012frontiers}.
 {QE and QSS methods require the knowledge of which species
 and reactions are fast. This knowledge can result from the
slow/fast decompositions performed numerically by CSP or ILDM
methods}, or from the calculation of a slowness index  \cite{radulescu2008robust}, in all cases
relying on trajectory simulation.
The application of these methods to computational biology is possible when model parameters are known.
When parameters are unknown, or if they are known only by their orders of magnitudes,
formal model reduction is needed. In addition, it is convenient to find reductions without
having to simulate trajectories.


 In this paper we propose a fully formal method to identify the slow and fast variables in a
 biochemical kinetic model with polynomial rate functions, without simulation of the trajectories.
 The method is based on computation of tropical
 equilibrations. Tropical methods \cite{litvinov2007maslov,maclagan2009introduction}, also known as {idempotent or} max-plus algebras due
 their name to the fact that one of the pioneer of the field, Imre Simon,
 was Brazilian. These methods found numerous applications to computer science \cite{simon1988recognizable},
 physics \cite{litvinov2007maslov}, railway traffic \cite{chang1998deterministic}, and  statistics \cite{pachter2004tropical}. We have shown recently that they can be used to analyse
 systems of polynomial or rational differential equations with applications
 to cell cycle modelling \cite{NGVR12sasb}.
 The main idea of our approach is to identify situations when two or several
 monomials of different signs equilibrate each other and dominate all the remaining
 monomials in the right hand side of the differential equations defining the chemical kinetics.
 We call this situation tropical equilibration  \cite{Noel2013a}. Tropical equilibration was previously used in an interesting
 study by Savageau \cite{savageau2009phenotypes} as a design tool for network steady states.
Our present focus is different because we are concerned with
dynamics and model reduction.
 We also propose an algorithm using the Newton polytope
 to solve the tropical equilibration problem efficiently for large biochemical networks.
  An alternative algorithm
 for finding  tropical equilibrations, based on constraint logic programming
 was proposed in \cite{soliman2014constraint}. However,
 when there are infinite branches of equilibrations, logic programming has no other alternative
but the exhaustive enumeration of solutions between arbitrary bounds,
whereas the Newton polytope method detects one solution per branch
which is enough for identifying variable timescales and reduced models.

\section{Approach}
{
\subsection{Dynamical equations and slow-fast decomposition} \label{slowfast0}
{The biochemical networks stemming from cell biology integrate processes evolving on very different
time scales. For instance, the changes of messenger RNA concentrations are usually  faster compared to
changes of protein concentrations and the post-transcriptional modifications of proteins
(for instance phosphorylation) are  faster than protein synthesis. For this reason, we  will
consider here slow-fast systems that have variables evolving on very different
timescales. Formally, variables $\vect{x}$ are much faster than variables $\vect{y}$ if the logarithmic
derivatives $\D{\log(\vect{x})}{t}$ are much larger in absolute values than $\D{\log(\vect{y})}{t}$.
After time rescaling, the differential equations describing
the dynamics of a system with fast variables $\vect{x}$ and slow variables $\vect{y}$ read as:}
\begin{eqnarray}
\D{\vect{x}}{t} & = \frac{1}{\eta} \vect{f}(\vect{x},\vect{y})  \label{fseq1} \\
\D{\vect{y}}{t} & =  \vect{g}(\vect{x},\vect{y}), \label{fseq2}
\end{eqnarray}
where {$\eta$ is a small positive parameter} and $\vect{f}$, $\vect{g}$ are functions not depending of $\eta$.

{In biochemical networks, the variables $\vect{x}$ and $\vect{y}$ are (positive) species concentrations.
Therefore, the functions $\vect{f}$, $\vect{g}$ are defined on the positive orthant.
Furthermore, for most of the kinetic laws, the
functions $\vect{f}$, $\vect{g}$ are polynomial or rational in the species concentrations. Although our methods apply
for both polynomial and rational functions, for the sake of simplicity we will consider that
$\vect{f}$ and $\vect{g}$ are polynomial functions. The system \eqref{fseq1},\eqref{fseq2} is endowed with positive
initial conditions for all variables:
\begin{equation}
\vect{x}(0)= \vect{x}_0 ,\, \vect{y}(0)= \vect{y}_0.
\label{initial}
\end{equation}
}

Let us suppose that the fast dynamics \eqref{fseq1} has a {unique} stable state $\vect{x}^*(\vect{y})$ for
all fixed $\vect{y}$ values. Let $\vect{J}(\vect{y})$ be the linear operator (Jacobian) that gives the linearization
of $\vect{f}(\vect{x},\vect{y})$ at fixed $\vect{y}$, namely
$$\vect{f}(\vect{x},\vect{y}) = \vect{J}(\vect{y}) (\vect{x} - \vect{x}^*(\vect{y})) + O(|\vect{x} - \vect{x}^*(\vect{y})|^2).$$
We say that the stable state $\vect{x}^*(\vect{y})$  is uniformly hyperbolic if all the eigenvalues
in the spectrum $Spec_{\vect{J}(\vect{y})}$
of $\vect{J}(\vect{y})$ have strictly negative real parts and are at a distance from the imaginary
axis larger than {a value} $d>0$, namely
\begin{equation}
{\text{ there is } d>0 \text{ such that }} Re(\lambda) < -d \text{ for all } \lambda \in Spec_{\vect{J}(\vect{y})} \text{ for all }\vect{y}.
\label{jacc}
\end{equation}
Tikhonov's theorem
\cite{tikhonov1952systems}
says that if the above conditions are satisfied, then after a quick transition the system
evolves approximately according to the following differential-algebraic equation:
\begin{eqnarray}
\D{\vect{y}}{t}  =  \vect{g}(\vect{x},\vect{y}) ,  \label{da1} \\
\vect{f}(\vect{x},\vect{y}) = 0.          \label{da2}
\end{eqnarray}
More precisely, the difference between solutions of  \eqref{fseq1},\eqref{fseq2}
and solutions of \eqref{da1},\eqref{da2} starting from the same
initial data satisfying \eqref{da2} {(i.e. $\vect{y}(0)= \vect{y}_0$,
$\vect{x}(0)= \vect{x}^*_0$, where $\vect{x}^*_0$ is the unique solution of
$\vect{f}(\vect{x},\vect{y}_0) = 0$)}
vanishes asymptotically like a positive power of $\eta$
when $\eta \to 0$.
{In the case when \eqref{fseq1} has several stable steady states,
then which one of these states will be chosen as solution of
\eqref{da2}
will depend on the
initial conditions \eqref{initial} of the full model.
}

Eq.~\eqref{da2} means that the fast variables are slaved by the slow ones. In this case, and given the
condition \eqref{jacc} on the Jacobian of $\vect{f}$  one can implicitly solve \eqref{da2} and transform \eqref{da1} into
an autonomous reduced model for the slow variables. This approach is known as quasi-steady state approximation.

The first and most important step in the implementation of this
reduction method is to find the slow-fast decomposition \eqref{fseq1},\eqref{fseq2}, which means to identify $\vect{x}$, $\vect{y}$ and $\eta$. For small models this can be done by rescaling variables and
kinetic constants and by identifying the small parameter $\eta$ as a ratio of kinetic
constants or initial values of the variables. A well known example is the quasi-steady state
approximation of the Michaelis-Menten enzymatic mechanism, when $\vect{x}$ is the concentration of the
enzyme-substrate complex, $\vect{y}$ is the substrate concentration and $\eta$ represents
the ratio of the enzyme to the substrate concentrations \cite{Noel2013a}. More generally, $\eta$
can be interpreted as the ratio of fast to slow timescales. Numerical methods such as
ILDM \cite{maas1992simplifying} use the Jacobian of the full system to obtain the {slow-fast decomposition.}
In such methods $\eta$ can be interpreted as the gap separating in logarithmic scale,
the timescales of slow and fast variables obtained from the spectrum of the Jacobian.
In this paper we present a symbolic method to perform the same decomposition. This method is
based on tropical geometry {\cite{litvinov2007maslov,maclagan2009introduction}}.
}

\subsection{Tropical equilibrations and timescales of the variables} \label{slowfast}
We consider biochemical networks described by the following differential equations
 \begin{equation}
 \D{x_i}{t} = \sum_{j \in [1,r]} k_j {S_{ij}}  \vect{x}^{\vect{\alpha_{j}}}, \, i \in [1,n].
 \label{massaction}
 \end{equation}
where $k_j >0, j \in [1,r]$ are kinetic constants, $r$ is the number of reactions, {$S_{ij}$ are the elements of the so-called stoichiometric matrix}, $\vect{\alpha}_{j} = (\alpha_1^j, \ldots, \alpha_n^j) \in \Z^n_+$ are multi-indices,
  $\vect{x^{\vect{\alpha}_{j}}}  = x_1^{\alpha_1^j} \ldots x_n^{\alpha_n^j}$ {and $x_i,\, i \in [1,n]$ are the species concentrations, $n$ being the number of species}.

{The polynomial equations \eqref{massaction} can result from the mass action law.
For instance, a reaction $A + B \to C$ of kinetic constant $k$ and satisfying the
mass action law, has $S_{11} = -1, S_{21} = -1, S_{31} = 1$, $\vect{\alpha}_1 = (1,1,0)$, which
correspond to the kinetic equations
 \begin{eqnarray}
 \D{x_1}{t} &= -k x_1 x_2, \notag \\
 \D{x_2}{t} &= -k x_1 x_2, \notag  \\
 \D{x_3}{t} &= k x_1 x_2,
 \end{eqnarray}
where $x_1$, $x_2$, $x_3$ are the concentrations of $A$, $B$, $C$, respectively.

 {We can notice} that the mass action law implies tight relations between $\alpha_{j}$ and $S_{ij}$, namely $\alpha_i^j = -S_{ij}$ if $S_{ij}<0$,
otherwise $\alpha_i^j = 0$. These relations are not {needed} in our approach.
{Furthermore, our method can be extended to the more general case when reaction
 rates are rational functions of the concentrations. Typically, we can use the least common denominator of
reaction rates to express the right hand sides of the kinetic equations as ratios of polynomials
and apply the method to the numerators. This extension was briefly discussed in \cite{NGVR12sasb}.}
}

In what follows, the kinetic parameters do not have to be known precisely {and}
they are given by their orders of magnitude.
{Usually, orders of magnitude are approximations of the parameters by integer
powers of ten and serve for rough comparisons.
Our definition of orders of magnitude is based on the equation}
\begin{equation}
k_j = \bar k_j \varepsilon^{\gamma_j}, \quad \gamma_j = \text{round}( \log(k_j) / \log(\varepsilon)),
\label{scaleparam}
\end{equation}
{where $\varepsilon$ is a positive parameter  smaller than $1$, $\gamma_j$ is the order of $k_j$,
$\bar k_j$ has order {zero} and round stands for the closest integer, {with half-integers rounded to even numbers}.
When $\epsilon = 1/10$, our definition provides the usual decimal orders.
Parameter order calculation is the first step of the algorithm in Sect.~\ref{pre-process}.}

 We must emphasize that the parameter $\varepsilon$ introduced in this section
 is not necessarily the fast/slow timescale ratio $\eta$ occurring in Tikhonov's theorem. As a matter of fact, as will be shown later in this section, the parameter $\eta$ can be expressed as a power of $\varepsilon$. In short, $\varepsilon$ is used just for expressing everything as powers.

From \eqref{scaleparam} it follows that if $\gamma_j \neq \gamma_i$ and $\gamma_i$ are integers,
then $k_i/k_j > 1/\varepsilon {>>1}$ or  $k_j/k_i > 1/\varepsilon {>> 1}$, meaning that the parameters $k_i$, $k_j$
are well separated.
However, the condition  $\gamma_j \neq \gamma_i$ is
not always needed in this approach.
All we need is the separation between the slow and fast
timescales resulting from our calculations {(this will be the gap condition introduced later in this section).
Networks with well separated constants were also studied in \cite{gorban-dynamic} for
the particular case of monomolecular reactions.}

Timescales of nonlinear systems depend not only on parameters but also on species concentrations, which are
a priori unknown. In order to compute them, we introduce a vector  $\vect{a} = (a_1,\ldots,a_n)$, such that
{\begin{equation}
\vect{x} = \bar{\vect{x}} \varepsilon^{\vect{a}}.
\label{aorders}
\end{equation}
The orders $a_i$ are generally rational and can be positive or negative. Of course, negative orders $a_i <0$ do
not mean negative concentrations, but very large concentrations, because  $\varepsilon < 1$. In general,
a higher $a_i$ means a smaller concentration $x_i$.}

Orders $\vect{a}$ are unknown and have to be calculated.
To this aim, the network dynamics can be described by a rescaled ODE system
 \begin{equation}
 \D{\bar{x}_i}{t} = (\sum_j \varepsilon^{\mu_j} \bar k_j {S_{ij}}   {\bar{\vect{x}}}^{\vect{\alpha_{j}}})\varepsilon^{-a_i},
 \label{massactionrescaled}
 \end{equation}
where
$\mu_j = \gamma_j +  \langle \vect{a},\vect{\alpha_j}\rangle,$
and $\langle , \rangle $ stands for the vector dot product. 

The r.h.s.\ of each equation in
\eqref{massactionrescaled} is a sum of multivariate monomials in the concentrations.
The exponents $\mu_j$ indicate how large are these monomials, in absolute value.
Generically, one monomial of exponent $\mu_j$ dominates the others $\mu_j < \mu_{j'}, j' \neq j$.
Accordingly, variables under the influence of a single dominant monomial, undergo
large changes in a short time.
The interesting case is when all variables are submitted to two dominant forces,
one positive and one negative and these forces have the same order.
We call this situation tropical equilibration (\cite{Noel2013a}).
More precisely, we have the following
\begin{definition}
We call tropical equilibration solutions the  vectors $\vect{a} \in \R^n$ {for which} the minimum in the definition of the piecewise-affine function
${\psi}_i(\vect{a}) = \min_j (\gamma_j + \langle \vect{a},\vect{\alpha_j} \rangle)$ is attained {for at least two} indices $j',j''$ corresponding to opposite signs monomials, i.e. $S_{ij'} S_{ij''}<0$.
{Equivalently, a tropical equilibration is a solution of the following system  of equations
for the orders $\vect{a}$:
\begin{equation}
\min_{j,S{ij}<0} (\gamma_j + \langle \vect{a},\vect{\alpha_j} \rangle) =
\min_{j',S{ij'}>0} (\gamma_j + \langle \vect{a},\vect{\alpha_j} \rangle),\quad i =1,\ldots,n .
\label{minplussystem}
\end{equation} }
\label{te}
\end{definition}
{For instance, if $\D{{x}_1}{t} =  \bar k_1  x_1  x_2 - \varepsilon^{1} \bar k_2  x_1 +
\varepsilon^{2} \bar k_3  x_2$, we have $\psi_1(\vect{a}) = \min \{ a_1+a_2, 1 + a_1, 2 + a_2  \}$.
Tropical equilibrations are solutions of
$\min \{ a_1+a_2, 2 + a_2 \} = 1 + a_1$, equivalent to
$a_1+a_2 = 1 + a_1 \leq 2 + a_2 $ or $2 + a_2 = 1 + a_1 \leq a_1+a_2$.
}

Intuitively, tropical equilibration means that dominant forces on variables compensate each other
and that variables change slowly under the influence
of the remaining weak forces. Compensation of dominant forces constrains
the dynamics of the system to a low dimensional invariant manifold \cite{NGVR12sasb,radulescu2012frontiers,Noel2013a}.

Tropical equilibrations are used to calculate the unknown orders $\vect{a}$ {as solutions of the system
\eqref{minplussystem}.
The solutions of \eqref{minplussystem} }have a
geometrical interpretation.
Let us define the extended order vectors $\vect{a}^e = (1, \vect{a}) \in \R^{n+1}$ and
extended exponent vectors $\vect{\alpha}_{j}^e= (\gamma_{j}, \vect{\alpha}_{j}) \in \Z^{n+1}$.
Let us consider the equality $\mu_j = \mu_{j'}$. This represents the equation of a
$n$ dimensional hyperplane of $\R^{n+1}$, orthogonal to the vector $\vect{\alpha_j}^e - \vect{\alpha_{j'}}^e$:
\begin{equation}
\langle \vect{a}^e,\vect{\alpha_j}^e \rangle = \langle \vect{a}^e,\vect{\alpha_{j'}}^e \rangle,
\label{lines}
\end{equation}
where $\langle,\rangle$ is the dot product in $\R^{n+1}$. We will see in the next section that
the minimality condition on the exponents $\mu_j$ implies that the normal vectors
$\vect{\alpha_j}^e - \vect{\alpha_{j'}}^e$
are edges of the so-called Newton polytope \cite{Henk2004,sturmfels2002solving}. {The algorithmic way to solve the set of
inequalities in Eq. \ref{minplussystem} along with the sign condition is described in Sects. \ref{newtonpoly}, \ref{pruning}. }

We call {\em tropically truncated system} the system obtained by pruning the system
\eqref{massactionrescaled}, i.e. by keeping only the dominating monomials.
{
 \begin{equation}
 \D{\bar{x}_i}{t} = \varepsilon^{\nu_i} (\sum_{j \in {D(i)}}  \bar k_j  S_{ij}  {\bar{\vect{x}}}^{\vect{\alpha_{j}}}),
 \label{massactionrescaledtruncated}
 \end{equation}
  where ${D(i)} = \underset{j}{\argmin} (\mu_{j},S_{ij}\neq 0)$ selects the dominating
 rates of reactions acting on species $i$ and
\begin{equation}
\nu_{i}  = \min \{ \mu_j |  S_{ij} \neq 0 \} - a_i .
\label{nui}
\end{equation}
 }

The tropically truncated equations contain generically two monomial
terms of opposite signs {(in special cases they can contain more than two terms
among which two have opposite signs)}. Polynomial systems with two monomial terms  are called {binomial or} toric. In systems biology, toric systems are known as S-systems and were used by Savageau \cite{savageau1987recasting} for modeling metabolic networks.

{The truncated system  \eqref{massactionrescaledtruncated} indicates
how fast is each variable, relatively to the others.
The {inverse} timescale of a variable $x_i$ is given by $ \frac{1}{x_i}\D{x_i}{t} = \frac{1}{\bar{x}_i}\D{\bar{x}_i}{t}$ that scales like $\varepsilon^{\nu_{i}}$.
Thus, if $\nu_{i'} < \nu_{i}$ then $x_{i'}$ is faster than $x_{i}$.

Let us assume that {$\nu_1 \leq \nu_2 \leq \ldots \leq \nu_n$
(this may require species re-indexing but is always possible) and} the following gap condition is fulfilled:
\begin{equation}
\text{there is } m < n \text{ such that } \nu_{m+1} - \nu_{m}  > 0,
\label{gapcondition}
\end{equation}
meaning that two groups of variables have separated timescales.
{The variables $\vect{X}_r = (x_1,x_2,\ldots,x_m)$ are fast (change significantly on
timescales of order of magnitude $\varepsilon^{-\nu_{m}}$ or shorter).
The remaining variables $\vect{X}_s = (x_{m+1},x_{m+2},\ldots,x_n)$ are
 slow (have little variation on timescales of order of magnitude $\varepsilon^{-\nu_{m}}$).
Then, the parameter $\eta = \varepsilon^{\nu_{m+1} - \nu_{m}}$ represents the fast/slow timescale
ratio in the Tikhonov's theorem from the preceding section.} Our gap condition means that
$\eta$ should be small. With these conditions, we have shown in \cite{Noel2013a,radulescu2015}
that quasi-steady state approximation can be applied. A further complication arises when the
system has fast cycles and this will be described in the next section. }

For systems with hierarchical relaxation, the separation between fast and slow variables is mobile
within the cascade of relaxing modes. {In the
extreme case this means that {all the species timescales are distinct and} separated by large enough gaps.}
Let us consider that we are interested in changes on timescales $\theta$ or slower. The timescale $\theta$
defines a threshold order value by the equation
\begin{equation}
\mu_{\text{threshold}}=-\log(\theta/\tau)/ \log(\varepsilon), \label{eq:mut_threshold}
\end{equation}
where $\tau$ are the time units from the model.
Then, from  \eqref{massactionrescaledtruncated} it follows that all variables {$x_i$} with $\nu_{i} \geq \mu_{\text{threshold}}$ are  slow. Perturbations in the concentrations of these species relax to an attractor slower or as slow as $\theta$.
The remaining species are fast and the perturbations in their concentrations relax to equilibrated values much
faster than $\theta$.

\subsection{Model reduction {of} fast cycles \label{sub:reduction}}
Tropical truncation is useful for identifying the slow and fast variables of a
system of polynomial differential equations. However, the truncation alone is not always enough for accurate reduction.
As discussed in \cite{Noel2013a,radulescu2015}, there are situations when the
truncated system is not a good approximation. Typically, truncation could eliminate all the reactions
exiting a fast cyclic subnetwork. Thus we get new conserved quantities, that were not conserved by the
full model. Truncation is in this case accurate at short times, but introduces
errors at large times.
In order to cope with fast cycles pruning, we adopt the recipe
discussed in \cite{GRZ10ces} for the quasi-equilibrium approximation. This recipe allows one to recover
the terms that were neglected by truncation, but which are important for large time dynamics.

{First, let us remind some definitions. We call linear conservation law of a system of differential equations,  a linear form $C(\vect{x}) = <\vect{c},\vect{x}> = c_1 x_1 + c_2 x_2 + \ldots + c_n x_n$
that is identically constant on trajectories of the system.
It can be easily checked that vectors in the left kernel $Ker^{l}(S)$ of the stoichiometric matrix $\vect{S}$
provide linear conservation laws of the system \eqref{massaction}. Indeed,
system \eqref{massaction} reads $\D{\vect{x}}{t} = \vect{S} \vect{R}(\vect{x})$,
where the components of the vector $\vect{R}$ are $R_j(\vect{x}) = k_j x^{\alpha_j}$. If
 $\vect{c} \vect{S} = 0$, then $\D{<\vect{c},\vect{x}>}{t} = \vect{c} \vect{S} \vect{R}(\vect{x}) = 0$, where $\vect{c} =(c_1,c_2,\ldots,c_n)$.

Let us assume that the truncated system  \eqref{massactionrescaledtruncated}, restricted
to the fast variables has a number of independent, linear conservation laws, defined by the
left kernel
vectors
$\vect{c}_1,\vect{c}_2,\ldots,\vect{c}_d$, where $\vect{c}_k = (c_{k1},c_{k2},\ldots,c_{kf})$.
These conservation laws can be calculated by recasting the truncated system
as the product of a new stoichiometric matrix and a vector of monomial rate
functions and further computing left kernel vectors of the new stoichiometric matrix.
We further assume that the fast conservation laws are not conserved by the
full system \eqref{massaction}.}


We define {the} new slow variables $\vect{Y}=(y_1,\ldots, y_d)$, where $y_k = \sum_{i=1}^f c_{ki} x_i$.
and eliminate the fast variables $x_1,x_2,\ldots,x_f$ by using the system :
\begin{eqnarray}
\sum_{j \in {D(i)}}   k_j  S_{ij}  {\vect{x}}^{\vect{\alpha_{j}}} &= &0,\, i \in [1,f] , \\
 \sum_{i=1}^f c_{ki} x_i & = &y_k, \, k \in [1,d] .
\end{eqnarray}
Reactions of the initial model that were pruned by truncation have to be restored if they act on the new slow variables $\vect{Y}$, i.e. if $\sum_{i=1}^f c_{li} S_{ik}    \neq 0$,
for some $l \in [1,d]$,
where $k$
is the index of the reaction to be tested.
Finally, the kinetic laws of these reactions have to be redefined in terms of the slow variables
$\vect{X}^s, \vect{Y}$.

{The rigourous justification of the reduction procedure for models
with fast cycles can be found in \cite{radulescu2015}.}

\section{Algorithm {to compute tropical equilibrations.}}
{In this section we introduce an algorithm allowing the automatic computation of tropical equilibrations. }
\subsection{Pre-processing}
\label{pre-process}
 We consider examples with polynomial vector field. The kinetic parameters of the equation system are scaled based on Eq.~\eqref{scaleparam}.
\subsection{Newton polytope and edge filtering}\label{edge-filtering}
\label{newtonpoly}
For each equation and species $i$, we define a Newton polytope {${\mathcal N}_i \subset \R^{n+1},$} that is the
convex hull
{of the union of all the half-lines  emanating in the positive $\epsilon$ direction
from the points $\alpha_j$ such that $S_{ij}\neq 0$ (thus, we first consider these
half-lines and then take their convex hull).}
This is the Newton polytope of the polynomial in right hand side of  Eq.~\eqref{massactionrescaled},
with the scaling parameter $\varepsilon$
considered as a new variable.
{If $\varepsilon$ does not appear in the coefficients of Eq.~\eqref{massactionrescaled},
then the half-lines above are replaced by the origins $\alpha_j$. The Newton
polytope is in this case the convex hull of the points $\alpha_j$ such
that $S_{ij}\neq 0$.}

\begin{figure}
\begin{center}
\includegraphics[width=0.94\textwidth]{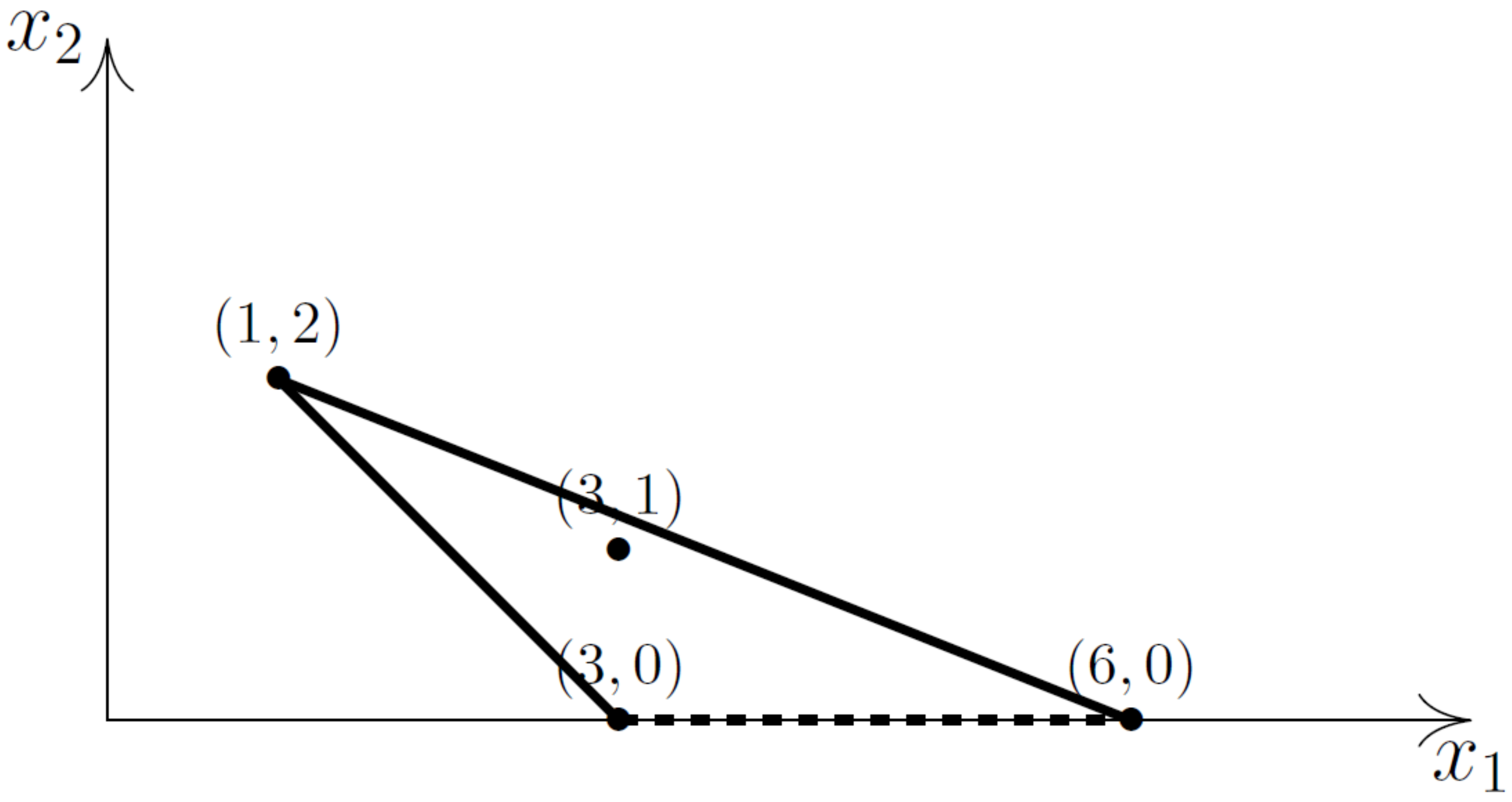}
\caption{ \label{fig:newton} \small
An example of a Newton polytope for the polynomial $-x_{1}^{6}+x_{1}^{3}x_{2}-x_{1}^{3}+x_{1}x_{2}^2$.
In this example, the monomial coefficients {do not depend on} $\varepsilon$ and
we want to solve
the tropical problem $\min(3a_1+a_2, a_1 + 2a_2) = \min(6a_1,3a_1)$.
The Newton polytope vertices $(6,0),(3,0),(1,2)$ are connected by lines.
The point $(3,1)$ is not a vertex as it lies in the interior
of the polytope. This stems to having $\min(3a_1+a_2, a_1 + 2a_2)=a_1+2a_2$
for all tropical solutions, which reduces the number of cases to be tested.
The thick edges satisfy the sign condition, whereas the dashed
edge does not satisfy this condition.
For this example, the solutions of the tropical problem are in infinite
number and are carried by the two
half-lines $a_1=a_2 \geq 0$ and $5a_1= 2a_2 \leq 0$, orthogonal to the
thick edges of the Newton polygon.
}

\end{center}
\end{figure}

As explained in Sect.~\ref{slowfast} the tropical equilibrations
correspond to vectors $\vect{a}^e = (1,\vect{a}) \in \R^{n+1}$
satisfying the optimality condition of Definition \ref{te}.
This condition is satisfied automatically on hyperplanes orthogonal to
edges of Newton polytope connecting vertices $\vect{\alpha}_{j'}^e$, $\vect{\alpha}_{j''}^e$
satisfying the opposite sign condition.
 {Therefore, a subset of edges from Newton polytope
is selected based on the filtering criteria which tells that the vertices belonging to an edge should be from opposite sign monomials as explained in Eq.~\eqref{eq:edge}.}
\begin{eqnarray}
E(P)=\{\{v_{1},v_{2}\}\subseteq\left(_{2}^{V}\right)\mid \text{conv}(v_{1},v_{2})\in F_{1}(P) , \nonumber \\
\wedge \, \text{sign}(v_{1})\times \text{sign}(v_{2})=-1\}, \label{eq:edge}
\end{eqnarray}
where $v_{i}$ is the vertex of the polytope and $V$ is the vertex
set of the polytope, $\text{conv}(v_{1},v_{2})$ is the convex hull of vertices
$v_{1},v_{2}$ and $F_{1}(P)$ is the set of 1-dimensional
face (edges) of the polytope.
 {$\text{sign}(v_{i})$ represents the sign of the monomial which corresponds to vertex $v_{i}$.

{Fig. \ref{fig:newton} shows an example of Newton polytope construction for a single equation
$\dot{x}_1 = -x_{1}^{6}+x_{1}^{3}x_{2}-x_{1}^{3}+x_{1}x_{2}^2$. Cf. definition \ref{te}
and eq.\eqref{minplussystem} tropical equilibrations are solutions of the equation
$$\min(3a_1+a_2, a_1 + 2a_2) = \min(6a_1,3a_1).$$
{Using a brute force method, tropical equilibration can be computed by solving the system}
\begin{eqnarray}
\text{case 1: } 3a_1+a_2 = 3a_1,\, 3a_1+a_2 \leq a_1 + 2a_2,\, 3a_1 \leq 6a_1 ,\label{c1} \\
\text{case 2: } 3a_1+a_2 = 6a_1,\, 3a_1+a_2 \leq a_1 + 2a_2,\, 3a_1 \geq 6a_1 ,\label{c2} \\
\text{case 3: } a_1+2a_2 = 3a_1,\, 3a_1+a_2 \geq a_1 + 2a_2,\, 3a_1 \leq 6a_1 ,\label{c3} \\
\text{case 4: } a_1+2a_2 = 6a_1,\, 3a_1+a_2 \geq a_1 + 2a_2,\, 3a_1 \geq 6a_1 .\label{c4}
\end{eqnarray}
The Newton polytope construction simplifies this task by automatically eliminating some
of the cases \eqref{c1},\eqref{c2},\eqref{c3},\eqref{c4}. We start by associating
a point in the $n$-dimensional space (here a plane, because there are two variables, $n=2$)
to each monomial of the polynomial equation. In Fig.~\ref{fig:newton} the points $(6,0)$,
$(3,1)$, $(3,0)$, $(1,2)$, correspond to the monomials
$x_{1}^{6}$, $x_{1}^{3}x_{2}$, $x_{1}^{3}$, $x_{1}x_{2}^2$, respectively.
The Newton polytope is the convex hull of the set of these points and has four vertices  (the point $(3,1)$ corresponding to the monomial $x_{1}^{3}x_{2}$ is internal to the polytope).
Each pair of points corresponding to monomials that have opposite signs in the original differential equation
indicates the choice of one of the four cases \eqref{c1},\eqref{c2},\eqref{c3},\eqref{c4}. For instance Case 1 means choosing the
pair of points $(3,1)$ and $(3,0)$. The Newton polytope construction allows one to identify the
cases involving internal points as redundant or impossible and to eliminate them.

Indeed, it can be easily checked that the cases 1 and 2 (Eqs.\eqref{c1},\eqref{c2})
have only the trivial solution $a_1=a_2=0$ that is also solution of cases 3 and 4.

Case 3 corresponds to the choice of the points $(1,2)$ and $(3,0)$ that are vertices of
the Newton polytope. The solution of \eqref{c3} is $a_1=a_2 \geq 0$ and describes a half line
orthogonal to the edge of the Newton polytope connecting the vertices $(1,2)$ and $(3,0)$.

Case 4 follows from the choice of the vertices $(1,2)$ and $(6,0)$. It has the solution
$2a_2 = 5 a_1 \le 0$ that describes a half line orthogonal to the corresponding
edge of the Newton polytope. }

Further definitions {and full proofs of the properties} of a Newton polytope can be found in
\cite{Henk2004,sturmfels2002solving}.}

\subsection{Pruning and feasible solutions} \label{pruning}
{We formalize here the pruning procedure illustrated for the simple example of the previous subsection.}

By feasible solution we understand a vector $(a_1,\ldots,a_n)$ satisfying
all the equations of the system \eqref{minplussystem}. A feasible solution lies in
the intersection of hyperplanes (or convex subsets of these hyperplanes) orthogonal
to edges of Newton polytopes obeying
the sign conditions. Of course, not all sequences of edges lead to
nonempty intersections and thus feasible solutions. This can be tested by the following
linear programming problem, resulting from  \eqref{minplussystem}:
\begin{equation}
\begin{split}
\gamma_j(i) + \langle \vect{a},\vect{\alpha_j(i)} \rangle =
\gamma_{j'}(i) + \langle \vect{a},\vect{\alpha_{j'}(i)} \rangle
\leq \gamma_{j''} + \langle \vect{a},\vect{\alpha_{j''}} \rangle),\\ \text{for all} \, j'' \neq j,j', {S_{ij''} \neq 0},\quad i =1,\ldots,n ,
\end{split}
\label{linprog}
\end{equation}
where $j(i),j'(i)$ define the chosen edge of the $i-$th Newton polytope.
The set of indices $j''$ can be restricted to vertices of the Newton polytope, because the inequalities
are automatically fulfilled for monomials that are internal to the Newton polytope.
For instance, for the example of the preceding section, the choice of the edge connecting vertices $(1,2)$ and $(3,0)$ leads to the following
linear programming problem:
 $$a_1 + 2a_2 = 3a_1 \leq 6a_1,\, {3a_1+a_2 \geq a_1 + 2a_2}, $$
 whose solution is a half-line orthogonal to the edge of the Newton polygon.

\begin{figure}
\begin{center}
\includegraphics[width=0.90\textwidth]{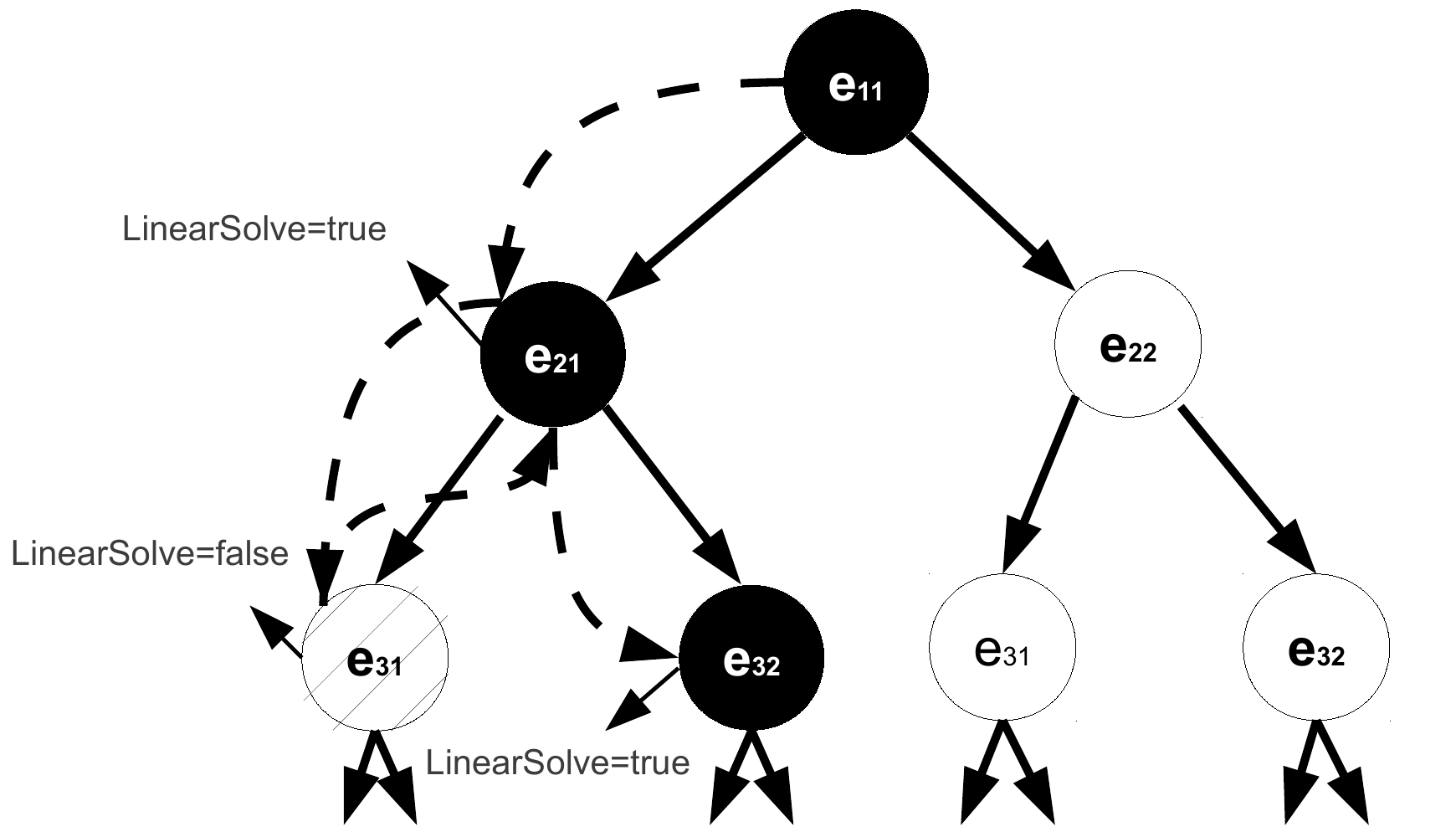}
\end{center}
\caption{\small \label{fig:pruning}Pruning strategy.
  {{This figure explains the pruning technique described in Sect.~\ref{pruning} and the evaluation of Eq.~\eqref{eq:cart}.}
The possible combinations of edges are represented in a tree representation. The algorithm starts
by testing for feasible solution for first
pair of edge sets. If a feasible solution is found, the algorithm
proceeds further to other edge sets or it backtracks. In the figure,
$e_{11}$ and $e_{21}$ are selected from edge sets $ne_{1}$, $ne_{2}$
and are checked for a feasible solution satisfying \eqref{linprog}. If such a solution exists, it moves to $e_{31}$from the
next edge set and again checks for feasible solution, if not then
it backtracks to $e_{21}$ and then to $e_{32}$ which results in a feasible
solution. Therefore, the sub-tree with root node $e_{31}$ is discarded
from future searches and this improves running time. Likewise the branch $e_{11}$ and $e_{22}$ is explored. This approach is similar to branch and bound algorithm technique.
The dashed arrows show the flow of the program.}}
\end{figure}

 {We introduce a pruning methodology (similar to a branch and bound algorithm technique) which helps to reduce the number of possible choices of Newton polytope edges leading to feasible solutions. Let us consider a system of polynomial
equations and order the equations as $eq_{1},eq_{2},...,eq_{n}$. Let the vertices of Newton polytope
${\mathcal N}_n$ be $v_{n1},v_{n2},...,v_{nl}$ where $l$ is the total number of vertices. The polytope edges are described by $ne_{1},ne_{2},...,ne_{n}$ where $ne_{i}$ denotes the set of edges from Newton polytope ${\mathcal N}_i$.
In order to search for feasible solutions
 an edge from each polytope needs to be selected. This translates to evaluating
the cartesian product of $ne_{1},ne_{2},\ldots, ne_{n}$ which can be
described by the following equation
\begin{eqnarray}
ne_{1}\times ne_{2}\times...\times ne_{n}=\{(e_{1j},e_{2j},...,e_{nj})\mid (e_{1j} \nonumber \\
\in ne_{1})\wedge (e_{2j}\in ne_{2})\wedge...\wedge (e_{nj}\in ne_{n})\}.\label{eq:cart}
\end{eqnarray}
where $e_{nj}$ is the $j^{th}$ edge in Newton polytope ${\mathcal N}_n$.
It is clear from the above that the possible choices are exponential.
In order to improve the running time of the algorithm, the pruning strategy evaluates Eq.~\eqref{eq:cart} in several steps(cf.
Algorithm \ref{alg:Tropical-Equilibration} and Fig. \ref{fig:pruning}). It starts with an arbitrary pair of edges and proceeds to add the next edge only when the inequalities \eqref{linprog} restricted to these two pair of edges are satisfied.
%
The corresponding set of inequalities can be solved using any standard linear programming package.}
\begin{algorithm}[htbp] \label{algo:1}
 \DontPrintSemicolon
\caption{SolveOrders: Steps of tropical equilibration algorithm\label{alg:Tropical-Equilibration}}
\BlankLine
\KwIn
{ {List of  edge sets $ne_{1},ne_{2},...,ne_{n}$, and the corresponding vertices}}
\KwOut{Orders of the variables $\vect{a}_{1},\vect{a}_{2},...,\vect{a}_{n}$
(tropical equilibration solution set)}
\Begin{
solutionset =\{\}; integer $k$=1; equation = \{\}
\BlankLine
SolveOrders(equation, $k$, edge-sets, vertices)
\BlankLine
\If {$k>n$}{ return}
\For {$l=1$ to number of entries in $ne_{k}$edge-set}{
equation(k)* = vertices in $l^{th}$row
\BlankLine
inequalities* = all other vertices in $ne_{1}$to $ne_{k}$edge-sets
\BlankLine
\If{ LinearSolve(equation,inequalities)is feasible} {
\If{ $k=n$ } {add the solution of LinearSolve to solutionset }
SolveOrders(equation, $k+1$,  $ne_{1},..,ne_{k}$, vertices)
}
}
}
*The equations and inequalities are initialised as per Eq.~\eqref{linprog}
\end{algorithm}

\subsection{Examples}
{As an illustration of our method we have chosen simple models that (i) have  polynomial dynamics and (ii) contain fast cycles that ask for the reduction steps described in Sect.\ref{sub:reduction}.
The Michaelis-Menten model of enzymatic reactions as well as
a cell cycle model proposed by Tyson \cite{tyson1991modeling} satisfy both these conditions.

{\subsubsection{The Michaelis-Menten model\label{sub:ExampleMM}}

The irreversible Michaelis-Menten kinetics consist of
three reactions:
\[
S + E \underset{k_{-1}}{ \overset{k_{1}}{\rightleftharpoons}} ES  \overset{k_2}{\rightarrow}
P + E,
\]
where $S,ES,E,P$ represent the substrate, the enzyme-substrate complex, the enzyme and the product,
respectively.

The corresponding system of polynomial differential equations reads:
\begin{equation}\label{full}
\begin{split}
 \dot{x}_1 & = -k_1 x_1 x_3 + k_{-1} x_2 ,\\
 \dot{x}_2 & = k_1 x_1 x_3 - (k_{-1}+k_2)x_2 ,\\
 \dot{x}_3 & = -k_1 x_1 x_3 + (k_{-1} +  k_2) x_2 ,\\
 \dot{x}_4 & = k_2 x_2 ,
\end{split}
\end{equation}
where $x_1=[S]$, $x_2=[ES]$, $x_3=[E]$, $x_4=[P]$.

The system \eqref{full} has two conservation laws $x_2 + x_3 = e_0$ and $x_1+x_2+x_4= s_0$. The
values $e_0$ and $s_0$  of the conservation laws result from the the initial conditions, namely
 $e_0 = x_2(0) + x_3(0)$ and $s_0 = x_1(0) +x_2(0) +x_4(0)$.

The conservation laws can be used to eliminate the variables $x_3$ and $x_4$ and obtain
the reduced system as follows
\begin{equation}\begin{split}
\label{mm}
\dot{x}_1 & = -k_1 x_1(e_0-x_2) + k_{-1} x_2 ,\\
\dot{x}_2 & = k_1 x_1 (e_0 - x_2) - (k_{-1}+k_2)x_2 .
\end{split}\end{equation}
There are two types of approximations and reductions for the Michaelis-Menten model, the quasi-steady
state and the quasi-equilibrium approximation \cite{meiske1978approximate,segel1988validity,segel1989quasi,GRZ10ces,gorban2011michaelis}. We discuss here how these
approximations can be related to tropical equilibrations (see also \cite{Noel2013a,soliman2014constraint} where the same model
is analysed using tropical curves).

Let us introduce orders of variables and parameters as follows  $x_i= \bar x_i \epsilon^{a_i}$, $1 \leq i \leq 2$,
$k_1= \bar k_1 \epsilon^{\gamma_1}$, $k_{-1}= \bar k_{-1} \epsilon^{\gamma_{-1}}$, $e_0= \bar e_0 \epsilon^{\gamma_e}$.

Then, we get the tropical equilibration equations by equating
minimal orders of positive monomials with minimal orders of negative monomials in \eqref{mm}:
\begin{align}
\gamma_{1}+ \gamma_{e} + a_1   &= \min (\gamma_{1} + a_1 ,\gamma_{-1} ) + a_2 ,
  \label{eq1} \\
\gamma_{1}+ \gamma_{e} + a_1  &= \min (\gamma_{1}+ a_1 , \min(\gamma_{-1},\gamma_2)  )+ a_2 .
 \label{eq2}
\end{align}

The quasi-equilibrium approximation corresponds to the case when
the reaction constant $k_{-1}$ is much faster than the reaction constant $k_2$. In
terms of orders, this condition reads $\gamma_{-1} < \gamma_2$.
In this case, the two tropical equilibration equations \eqref{eq1}, \eqref{eq2} are identical,
because $\min(\gamma_{-1},\gamma_2) = \gamma_{-1}$.
Let $\gamma_m = \gamma_{-1} - \gamma_1$ denote the order of the parameter $K_m = k_{-1}/k_1$.
There are two branches of solutions of \eqref{eq1}, namely
$a_2 = \gamma_{e}, a_1 \leq \gamma_m $ and
$a_2 = a_1 + \gamma_{e} - \gamma_{m}, a_1 \geq \gamma_m$
corresponding
to
 $\min (\gamma_{1} + a_1 ,\gamma_{-1} ) = \gamma_{1} + a_1 $ and to
 $\min (\gamma_{1} + a_1 ,\gamma_{-1} ) = \gamma_{-1}$, respectively.
Using the relation between orders and concentrations we identify the first branch of solutions with
the saturation regime $x_2 \approx e_0$ (the free enzyme is negligible) and
$x_1 >> K_m$ (the substrate has large concentration)
and the second branch with the linear regime $x_2 << e_0$ (the
concentration of the
attached enzyme is negligible) and $x_1 << K_m$ (the substrate has low concentration).

 In the linear regime of quasi-equilibrium the fast truncated system (obtained after removing all dominated monomials from \eqref{mm}) reads
\begin{equation}\begin{split}
\label{mmtrunc}
\dot{x}_1 & = -k_1 x_1 e_0  + k_{-1} x_2 , \\
\dot{x}_2 & =  k_1 x_1 e_0  - k_{-1}x_2 .
\end{split}\end{equation}
The variable $y= x_1 + x_2$ is conserved by the fast truncated system \eqref{mmtrunc}, but
not by the full system \eqref{mm}. Therefore, $y$ has to be considered as a new
slow variable.
By summing the two equations of \eqref{mm} term by term
we get
\begin{equation}
\dot{y} = - k_2 x_2 .
\label{eq:y}
\end{equation}

Using the quasi-equilibrium equation $-k_1 x_1 e_0  + k_{-1} x_2 = 0$
we eliminate $x_1$, $x_2$ by expressing them as $x_1 = y /(1+ k_1 e_0/k_{-1})$, $x_2 = y/(1 + k_{-1}/(k_1 e_0))$. Finally, we get the reduced model for the slow variable $y$,
\begin{equation}
\dot{y} = - k_2 y / (1 + k_{-1}/(k_1 e_0) ) = - V_{max} y /(e_0 + K_m) ,
\end{equation}
where $V_{max} = k_2 e_0$.

If we express
$\dot{y}$ as a function of the substrate concentration $x_1$ we obtain
$\dot{y} = - (V_{max}/K_m) x_1$, which is the well known Michaelis-Menten reaction
rate in the linear regime.

In the saturated quasi-equilibrium regime, the fast truncated system reads
\begin{equation}\begin{split}
\label{mmtruncs}
\dot{x}_1 & = -k_1 x_1(e_0-x_2), \\
\dot{x}_2 & =  k_1 x_1(e_0-x_2).
\end{split}\end{equation}

From \eqref{mmtruncs} we get the quasi-equilibrium equation $x_2 = e_0$ and further, using \eqref{eq:y}, we find the reduced model
\begin{equation}
\dot{y} = - V_{max}.
\end{equation}

The tropical method also allows us to test that variables $x_1$, $x_2$ are faster than $y$, which means that the reductions are consistent (fast variables are eliminated and the reduced
model is written in the slow variables only).
In terms of $\nu$ orders defined by eq.\eqref{nui},
one has to check that $\nu_1 < \nu_y$ and $\nu_2 < \nu_y$.
Using eq.\eqref{nui} together with the quasi-equilibrium condition,
we find that $\nu_2=\gamma_{-1}$ in the linear regime and
$\nu_2=\gamma_{1} + a_1$ in the saturated regime. Furthermore,
$\nu_1 \leq \gamma_{-1} + a_2 - a_1$,
 $\nu_y = \gamma_2 + a_2 - \min (a_1,a_2)$ for both regimes. The condition  $\nu_1 < \nu_y$
 is satisfied because $\gamma_2 > \gamma_{-1}$. $\nu_2 < \nu_y$
is satisfied in the linear regime because $\gamma_2 > \gamma_{-1}$. The same condition is
satisfied also in the saturated regime because $a_1 \leq \gamma_m = \gamma_{-1}-\gamma_{1}$ in this regime.

To summarize, the unique condition for quasi-equilibrium is $\gamma_2 > \gamma_{-1}$.
In particular, this approximation does not depend on the initial data
because $\gamma_e$ does not occur in the above condition.

The quasi-steady state approximation corresponds to the situation when $x_2$ is equilibrated
and  faster than $x_1$. In this case one has to combine
 \eqref{eq2} with the condition $\nu_2 < \nu_1$.
 Let us denote by $\gamma_m = \min(\gamma_{-1},\gamma_2) - \gamma_1$ the order of the
 parameter $K_m = (k_{-1} + k_2)/ k_1$.
 Eq.\eqref{eq2} alone has two branches of solutions. The first branch
 is defined by $a_1 \leq \gamma_m$, $a_2 = \gamma_e$ and corresponds to the saturated
 regime of quasi-steady state. The second branch is defined by
  $a_1 \geq \gamma_m$, $a_2 = a_1 + \gamma_e - \gamma_m$ and corresponds to the linear
  regime.
  From \eqref{mm} we find $\nu_1 = \min (\gamma_1 + a_1 + \gamma_e, \gamma_1+a_1+a_2,\gamma_{-1}+a_2) - a_1$ and $\nu_2 = \gamma_1 + a_1 + \gamma_e - a_2$.
  By elementary inequality algebra it follows that the condition $\nu_2 < \nu_1$
  is equivalent to $a_1 < \gamma_e$ at
  saturation and to $\gamma_m < \gamma_e$ in the linear regime.

  Summarizing, the conditions for quasi-steady state are
  $a_1 < \min (\gamma_m,\gamma_e)$ (saturation)
  or $\gamma_m < \min  (a_1, \gamma_e)$ (linear regime).
  In contrast to quasi-equilibrium, quasi-steady state depends on the initial conditions.

The quasi-steady state equations at saturation are $k_1  x_1 (e_0 - x_2)=0$, leading to
$x_2 = e_0$. In the linear regime one has $k_1 x_1 e_0 - (k_{-1} + k_2) x_2 =0$, leading
to $x_2 = e_0 x_1 / K_m$. Using \eqref{eq:y} we get the well known expressions  $\dot{y} = -k_2 e_0 = -V_{max}$ and $\dot{y} = - V_{max} x_1 /K_m$
representing the reaction rate in the saturated and linear regimes, respectively.

The timescales of variables and the validity of quasi-steady state
for Mi\-chaelis-Menten irreversible kinetics were previously derived by Segel \cite{segel1988validity,segel1989quasi}. Our time scales and conditions are
compatible with the ones of Segel on pieces, i.e. in
the linear and in the saturated regime of quasi-steady state.
For instance, like in \cite{segel1988validity} our conditions imply
that quasi-steady state can be valid for small $\gamma_e$ (large enzyme)
provided that $\gamma_m$ is smaller (very large $K_m$).

}

\subsubsection{The cell cycle model\label{sub:Example}}
This model describes the interaction between cyclin and cyclin-dependent kinase cdc2
during the progression of the eukaryotic cell cycle (see Fig.~\ref{fig:cellcyclemodel}).
Cyclin (variable $x_5$) is synthesized during interphase stage of the cycle
(reaction of constant $k_6$). Newly synthesized cyclin forms a complex with the phosphorylated kinase cdc2 (cdc2 is the variable $x_2$ and the complex formation reaction has constant $k_4$). The resulting complex (variable $x_4$) is called inactive or pre-maturation
promoter factor (pre-MPF). pre-MPF needs to be activated for enter into mitosis
in order to phosphorylate many substrates controlling
processes essential for nuclear and cellular division. The active form of MPF (variable $x_3$)
is produced from pre-MPF either by a non-regulated transformation (reaction of constant $k_{10}$) or by
an autocatalytic process (reaction of constant $k_9$). At the end of mitosis the active complex
dissociates (reaction of constant $k_1$), resulting in the phosphorylated cyclin (variable $x_6$)
that is degraded (reaction of constant $k_8$) and the de-phosphorylated kinase cdc2 (variable $x_1$).
The kinase is equilibrated with its phosphorylated form (variable $x_2$) by phosphorylation and
dephosphorylation reactions (of constants $k_2$ and $k_3$ respectively).

The full model has a stable periodic attractor, a limit cycle. The stable limit cycle oscillations
correspond to the periodic succession of interphase and mitosis phases of the cell cycle.
}

The corresponding system of differential equations
along with conservation laws
for the above model can be described as
\begin{eqnarray}
	 	\dot{x}_1 &=& k_1 x_3  - k_2  x_1  + k_3  x_2, \,
	 	\dot{x}_2 = k_2  x_1  - k_3  x_2  - k_4  x_2  x_5,  \notag \\
	 	\dot{x}_3 &=& k_{10}  x_4  - k_1  x_3  + k_9  x_3^2 x_4, \notag \\
	 	\dot{x}_4 &=& k_4  x_2  x_5  - k_{10}  x_4  - k_9  x_3^2 x_4, \,
	 	\dot{x}_5 = k_6  - k_4  x_2  x_5,\notag \\
	 	\dot{x}_6 &=& k_1  x_3  - k_8  x_6,   \,
        x_1 +  x_2 +  x_3 +  x_4 = 1.
\label{tyson67}
\end{eqnarray}
 {The value of the conservation law $x_1 +  x_2 +  x_3 +  x_4 = 1$ follows
 from the initial conditions $\vect{x}(0)=(0,0.75,0,0.25,0,0)$
 that were taken from \cite{tyson1991modeling}. Other initial conditions
 with the same value of the conservation law would lead to the same
 tropical equilibration solutions.

 Applying Definition \ref{te} and eq.\eqref{minplussystem} to this model, we obtain
 the following tropical equilibration problem:}
 \begin{eqnarray}
 \min(a_3 + \gamma_1,a_2+ \gamma_3) = a_1 + \gamma_2, & \notag \\
 a_1 + \gamma_2 = \min(a_2 + \gamma_3,a_2 + a_5 + \gamma_4), &  \notag \\
 \min(a_4 + \gamma_{10},2a_3 + a_4 + \gamma_9 ) = a_3 + \gamma_1, & \notag \\
 a_2 + a_5 + \gamma_4= \min(a_4 + \gamma_{10},2a_3 + a_4 + \gamma_9 ), &\notag \\
 \gamma_6=a_2 + a_5 + \gamma_4, & \notag \\
a_3 + \gamma_1 = a_6 + \gamma_8,
 \min(a_1,a_2,a_3,a_4)= 0 .&
\end{eqnarray}
Using the numerical values of the parameters from the original paper we find, for $\varepsilon = 1/9$,
$\gamma_{1}=0$, $\gamma_{2}=-6$, $\gamma_{3}=-3$, $\gamma_{4}=-2$, $\gamma_{6}=2$,
$\gamma_{8}=0$, $\gamma_{9}=-2$, $\gamma_{10}=2$ {(cf. Eq. \ref{scaleparam} and Sect. \ref{pre-process}).}

{\em Remark:} One may notice that the orders $\gamma$ depend on which units were used for the parameters. However,
if the parameter units are changed, the set of tropical equilibrations is transformed into
an equivalent one. Indeed, the model equations should be invariant with respect to units conversion.
In particular, if units of second  order reaction constants (i.e. coefficients of second order monomial rates)
are multiplied by $k$, one should subtract $log(k)/log(\varepsilon)$ from the parameter orders and add
the same quantity to the concentration orders. This will generate an equivalent set of solutions,
up to {rounding} errors.

Using our algorithm {(cf. Sects.\ref{newtonpoly}, \ref{pruning})} we got three tropical equilibrations for this system, namely
$\vect{a}_1=(8,5,2,0,-1,2)$,
$\vect{a}_2=(5,2,2,0,2,2)$,
$\vect{a}_3=(3,0,2,0,4,2)$.

The rescaled truncated system for the solution $\vect{a}_3$ reads
\begin{eqnarray}
\dot{\bar x}_1 =\varepsilon^{-6}(\bar k_3 \bar x_2 - \bar k_2 \bar x_1),
\dot{\bar x}_2 =\varepsilon^{-3}(\bar k_2 \bar x_1 - \bar k_3 \bar x_2), & \notag \\
\dot{\bar x}_3 =\bar k_{10} \bar x_4 - \bar k_1 \bar x_3 + \bar k_9 \bar x_3^2 \bar x_4, & \notag \\
\dot{\bar x}_4 = \varepsilon^2(-\bar k_{10}  \bar x_4 + \bar k_4 \bar x_2 \bar x_5 - \bar k_9 \bar x_3^2 \bar x_4), & \notag \\
\dot{\bar x}_5 = \varepsilon^{-2}(\bar k_6 -\bar k_4 \bar x_2 \bar x_5 ),
\dot{\bar x}_6 = \bar k_1 \bar x_3 - \bar k_8 \bar x_6.
\end{eqnarray}
It appears clearly that the variables $x_1$, $x_2$, $x_5$ are fast.
{More precisely, their characteristic times are
 $\nu_1^{-1} = \varepsilon^{6}$,  $\nu_2^{-1} = \varepsilon^{3}$,
 $\nu_5^{-1} = \varepsilon^{2}$, respectively. The largest of these timescales
is here approximately $0.01$ (in minutes which are the time units of the model).
The remaining slow variables have characteristic times from  $\varepsilon^0$ to $\varepsilon^{-2}$,
i.e. approximately from $1$ to $100$ min.
Therefore, the timescales of slow and fast species are separated by a gap, and the
singular perturbation small parameter (cf. Sect.\ref{slowfast0}) is
$\eta = t_{fast}/t_{slow} \sim \varepsilon^2$ (the power 2 arises as the
difference between $\nu_6 = \nu_3 = 0$, coming from the
fastest slow species and $\nu_5 = -2$, coming from the slowest fast species).
}

The
fast truncated system reads
\begin{eqnarray}
\dot{x}_1=k_3 x_2 - k_2 x_1,
\dot{x}_2=k_2 x_1 - k_3 x_2, & \notag \\
\dot{x}_5= k_6 - k_4 x_2 x_5.
\label{ft}
\end{eqnarray}
and has a single conservation law $C_1 = x_1 + x_2$ that provides a new
slow variable.
{This conservation law, not conserved by the full system \eqref{tyson67}, indicates the presence of a fast cycle in the model. It is the rapid phosphorylation/dephosphorylation cycle transforming the cyclin $x_1$ into its phosphorylated form $x_2$ and back. }
 The fast variables are eliminated from the system obtained by adding to
\eqref{ft} the definition of the fast conservation law {cf. Sect.\ref{sub:reduction}:}
\begin{equation}
k_3 x_2 - k_2 x_1 = 0, k_6 -k_4 x_2 x_5 = 0, y = x_1 + x_2.
\label{eq:invman}
\end{equation}

The differential equation for $y$ is obtained by adding the first two equations of the full system
\eqref{tyson67}, and thus restoring the terms $k_1  x_3$ and $k_4  x_2  x_5$, that have
order $\varepsilon^2$ and were pruned in the first step.

Finally, we obtain the following reduced model
\begin{eqnarray}
\dot{x}_3= k_{10} x_4 - k_1 x_3 + k_9 x_3^2x_4, & \label{eqred1} \\
\dot{x}_4= -k_{10}  x_4 + k_6 - k_9 x_3^2x_4, & \label{eqred2} \\
\dot{x}_6 = k_1 x_3 - k_8 x_6, \dot{y} = k_1 x_3 - k_6. \label{eqred3}
\end{eqnarray}
and the slaved fast variables are given by $x_1 = y k_2/(k_2+k_3) \approx y k_2/k_3 $, $x_2 = y k_3/(k_2+k_3) \approx y$, $x_5 = k_6 (k_2 + k_3)/ (k_4 k_3 y) {\approx
k_6 / (k_4  y)}$, where we have used {\eqref{eq:invman}
and}
the fact that $k_2 \ll k_3$.

{Let us note that the variable $y$ has the same order as $x_2$ ($a_y = a_3 = 2$), it is tropically equilibrated
($\gamma_1+a_3 = \gamma_6 = 2$ in Eq.\eqref{eqred3}), and has $\nu_y = \gamma_6  - a_y =0$ meaning that it is slow. }

Let us call this four variables model reduced model 1. Note that in this model the dynamics of the
variables $x_3,x_4$ is decoupled from the two others. We can therefore
conclude that by our approach we obtain a two dimensional
minimal cell cycle model.

Repeating the procedure for the equilibrations $\vect{a}_1$, $\vect{a}_2$ we find two other rescaled
truncated systems.

The rescaled truncated system for the solution $\vect{a}_1$ reads
\begin{eqnarray}
\dot{\bar x}_1=\varepsilon^{-6}(\bar k_3 \bar x_2 - \bar k_2 \bar x_1 + k_1 x_3),
\dot{\bar x}_2=\varepsilon^{-3}(\bar k_2 \bar x_1 - \bar k_3 \bar x_2), & \notag \\
\dot{\bar x}_3=\bar k_{10} \bar x_4 - \bar k_1 \bar x_3 + \bar k_9 \bar x_3^2 \bar x_4, & \notag \\
\dot{\bar x}_4= \varepsilon^2(-\bar k_{10}  \bar x_4 + \bar k_4 \bar x_2 \bar x_5 - \bar k_9 \bar x_3^2 \bar x_4), & \notag \\
\dot{\bar x}_5= \varepsilon^{3}(\bar k_6 -\bar k_4 \bar x_2 \bar x_5 ),
\dot{\bar x}_6= \bar k_1 \bar x_3 - \bar k_8 \bar x_6,
\end{eqnarray}
and for the solution $\vect{a}_2$ we got
\begin{eqnarray}
\dot{\bar x}_1=\varepsilon^{-6}(\bar k_3 \bar x_2 - \bar k_2 \bar x_1),
\dot{\bar x}_2=\varepsilon^{-3}(\bar k_2 \bar x_1 - \bar k_3 \bar x_2), & \notag \\
\dot{\bar x}_3=\bar k_{10} \bar x_4 - \bar k_1 \bar x_3 + \bar k_9 \bar x_3^2 \bar x_4, & \notag \\
\dot{\bar x}_4= \varepsilon^2(-\bar k_{10}  \bar x_4 + \bar k_4 \bar x_2 \bar x_5 - \bar k_9 \bar x_3^2 \bar x_4), & \notag \\
\dot{\bar x}_5= (\bar k_6 -\bar k_4 \bar x_2 \bar x_5 ),
\dot{\bar x}_6= \bar k_1 \bar x_3 - \bar k_8 \bar x_6.
\end{eqnarray}

In both cases, the variable $x_5$ is slow, {which was not the case for the
equilibration $\vect{a}_1$. This is possible, because for a nonlinear model, the
timescale of a variable depends on the concentration range in which the model functions.
The equilibrations $\vect{a}_1$ and $\vect{a}_2$ correspond to {very low and low} concentrations of phosphorylated
kinase $x_2$ (proportional to $\varepsilon^5$ and $\varepsilon^2$, respectively), meaning slow consumption of the cyclin $x_5$. The concentration of $x_2$ is large for the equilibration $\vect{a}_3$ (proportional to $\varepsilon^0$) leading to rapid consumption of $x_5$ (see Eq.\eqref{tyson67}).}

The two tropical equilibrations {$\vect{a}_1$ and $\vect{a}_2$} lead to the same reduced model,
which we call reduced model 2:
\begin{eqnarray}
\dot{x}_3= k_{10} x_4 - k_1 x_3 + k_9 x_3^2x4, & \notag \\
\dot{x}_4= -k_{10}  x_4 + k_6 - k_9 x_3^2x_4, & \notag \\
\dot{x}_5 = k_6  - k_4  y  x_5, & \notag \\
\dot{x}_6= k_1 x_3 - k_8 x_6, \dot{y} = k_1 x_3 - k_6,
\end{eqnarray}
 to be considered together with  $x_1 =  y k_2/k_3 $, $x_2 =  y$.

The tropical setting confirms ideas from the theory of nonlinear dynamical systems. The
two reduced models are nested.
Reduced model 2 has a larger number of slow (relaxing) variables than reduced model 1.
This means that the corresponding invariant manifolds are embedded one into another with
the lowest dimensional one defined by the reduced model 1 carrying the dynamics on the
limit {cycle} attractor. {Starting with initial low concentrations of the phosphorylated kinase
corresponding to the equilibration $\vect{a}_1$ or $\vect{a}_2$, the system will increase these concentrations
to levels corresponding to the equilibration $\vect{a}_3$ that allow the stable
limit cycle oscillations.}

%

{
One can notice that our reduced model 1 does not contain the parameters
$k_2$, $k_3$, $k_4$ of the full model. This means that as long as the
phosphorylation and the dephosphorylation of the free kinase, as well
as the formation of cyclin kinase complex are fast enough, the actual
values of the kinetic constants of these processes are not important.


In his paper, Tyson \cite{tyson1991modeling} also proposes a two variables reduced model:
\begin{eqnarray}
\dot{u} &=& k_9 (v - u) (\alpha +u^2) - k_1 u, \notag \\
\dot{v} &=& k_6 - k_1 u, \label{reducedtyson}
\end{eqnarray}
where $u=x_3$, $v=x_3+x_4+x_5$, $\alpha =k_{10}/k_{9}$.

It can be easily checked that Eqs.\eqref{reducedtyson}
are equivalent with our Eqs.\eqref{eqred1},\eqref{eqred2}, provided that
$x_5 \ll x_4$ and $x_5 \ll x_3$. These last conditions, justified by intuitive arguments,
were used in the derivation of the reduced model in \cite{tyson1991modeling}.
In our approach, the same conditions follow immediately
from the orders of the species concentrations. Indeed, for the equilibration $\vect{a}_3$
we have $x_5 \sim \varepsilon^4$, $x_3 \sim \varepsilon^2$, $x_4 \sim \varepsilon^0$, therefore
$x_5 \ll x_4$ and $x_5 \ll x_3$.
To summarize, the advantage of our approach is that it is automatic
and can be applied to larger models that are more difficult or impossible
to grasp by simple intuition.
}


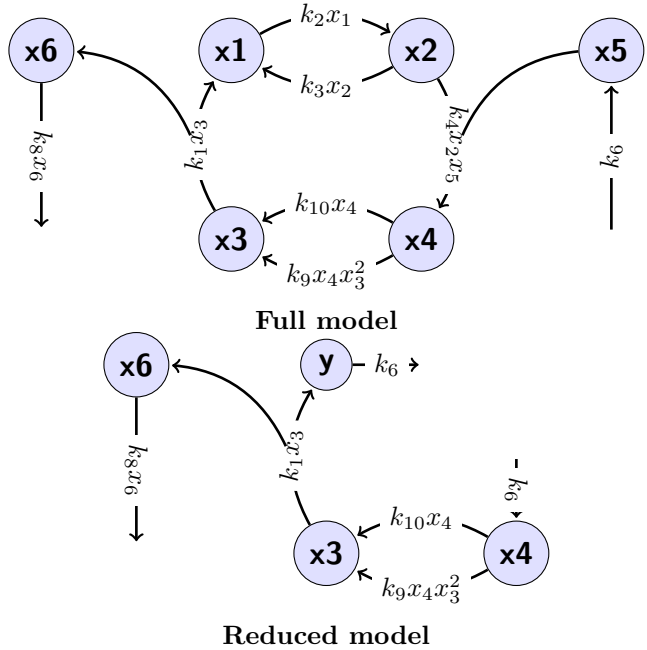
\begin{figure}
\begin{center}
\scalebox{1}{
\begin{tikzpicture}
 \SetUpEdge[lw         = 0.5pt,
            color      = black,
            labelstyle = {fill=white, sloped}]
  \tikzset{node distance = 2.5cm,main node/.style={circle,fill=blue!12,draw,font=\sffamily\large\bfseries}
   }
 \GraphInit[vstyle= normal]
  \SetGraphUnit{2}
\node[main node] (x1){x1};
\node[main node] (x2)[right of=x1]{x2};
\node[main node] (x6)[left of=x1] {x6};
\node[main node] (x5)[right of=x2] {x5};
\node [main node](x3)[below of=x1] {x3};
\node [main node](x4)[below of=x2] {x4};
\node [](x6b)[below of=x6] {};
\node [](x5b)[below of=x5] {};
 \tikzset{EdgeStyle/.style={post,line width = 1}}
  \tikzstyle{EdgeStyle}=[post,bend left,line width = 1]
 \Edge[label=$k_{4}x_2 x_5$](x2)(x4)

 \Edge[label=$k_1 x_3$](x3)(x1)
 \Edge[label=$k_2x_1$](x1)(x2)
 \tikzstyle{EdgeStyle}=[post,bend left,line width = 1]
  \Edge[label=$k_3x_2$](x2)(x1)
  \Edge[label=$k_9 x_4 x_3^2$](x4)(x3)
   \tikzstyle{EdgeStyle}=[post,bend right,line width = 1]
  \Edge[label=$k_{10}x_4$](x4)(x3)
\path[] (x2) to [bend left] node[midway](O){} (x4) ;
\path[] (x3) to [bend left] node[midway](O2){} (x1) ;
  \tikzstyle{EdgeStyle}=[bend right,line width = 1]
  \Edge[](x5)(O)
   \tikzstyle{EdgeStyle}=[post,bend right,line width = 1]
  \Edge[](O2)(x6)
  \tikzstyle{EdgeStyle}=[post,line width = 1]
  \Edge[label=$k_{8}x_6$](x6)(x6b)
  \Edge[label=$k_{6}$](x5b)(x5)
\end{tikzpicture}
}
\centerline{\bf Full model}
\scalebox{1}{
\begin{tikzpicture}
 \SetUpEdge[lw         = 0.5pt,
            color      = black,
            labelstyle = {fill=white, sloped}]
  \tikzset{node distance = 2.5cm,main node/.style={circle,fill=blue!12,draw,font=\sffamily\large\bfseries}
   }
 \GraphInit[vstyle= normal]
  \SetGraphUnit{2}
  \node [main node](x3){x3};
  \node [main node](x4)[right of=x3]{x4};
\node[main node] (x1)[above of=x3]{y};
\node[main node] (x6)[left of=x1] {x6};
\node [](x6b)[below of=x6] {};
\node [](x2)[right of=x1] {};
\coordinate (O1) at ($(x1)!0.5!(x2)$) ;
\coordinate (O2) at ($(x2)!0.5!(x4)$) ;
\path[] (x3) to [bend left] node[midway](O3){} (x1) ;
 \tikzset{EdgeStyle/.style={post,line width = 1}}
  \tikzstyle{EdgeStyle}=[post,bend left,line width = 1]
 \Edge[label=$k_1 x_3$](x3)(x1)
 \tikzstyle{EdgeStyle}=[post,bend left,line width = 1]
  \Edge[label=$k_9 x_4 x_3^2$](x4)(x3)
   \tikzstyle{EdgeStyle}=[post,bend right,line width = 1]
  \Edge[label=$k_{10}x_4$](x4)(x3)
  \Edge[](O3)(x6)
  \tikzstyle{EdgeStyle}=[post,line width = 1]
  \Edge[label=$k_{8}x_6$](x6)(x6b)
  \Edge[label=$k_{6}$](x1)(O1)
   \tikzstyle{EdgeStyle}=[post,line width = 1]
  \Edge[label=$k_{6}$](O2)(x4)
\end{tikzpicture}
}
\centerline{\bf Reduced model}
\end{center}
\caption{Graphic representation of the
full cell cycle model \cite{tyson1991modeling} (referenced as BIOMD00000005 by Biomodels)
and of the reduced model 1. The full
model describes the cyclin production and complex formation
between the cyclin
and the kinase cdc2, the autocatalytic activation
(by dephosphorylation), and the dissociation of this complex
followed by the destruction of the cyclin.
The reduced model represents accurately the
same processes, on the invariant manifold containing
the periodic attractor.
The different variables mean:
$x_1: \, \text{cdc2}$,
$x_2: \, \text{cdc2-P}$,
$x_3: \, \text{cdc2}:\text{cyclin-P}$ i.e. active
MPF complex,
$x_4: \, \text{P-cdc2}:\text{cyclin-P}$ i.e. pre-MPF
complex,
$x_5: \, \text{cyclin}$,
 $x_6: \, \text{cyclin-P}$,
  $y  = x_1 + x_2: \, \text{total free cdc2}$. Both full and reduced model
  are biochemical networks with polynomial rate functions.
}
\label{fig:cellcyclemodel}
\end{figure}

\section{Results }
\subsection{Details on the implementation of the algorithm}
The differential equation system along with the conservation laws
for a given biochemical model (in SBML format) are generated using
the pocab software\cite{Samal2012}.  Polymake \cite{Gawrilow2000} is used to compute the Newton
polytope from the set of exponent vectors. For solving the linear programming we used Gurobi\cite{optimization2012gurobi} in Java programming environment. The expressions in the equation system are processed using computer algebra system Maple.
\footnote{The code can be downloaded together with supporting information from  {http://www.abi.bit.uni-bonn.de/index.php?id=17.}}

\subsection{Results on  Biomodels database} \label{results}
\begin{figure}
\begin{center}
\includegraphics[width=0.5\textwidth]{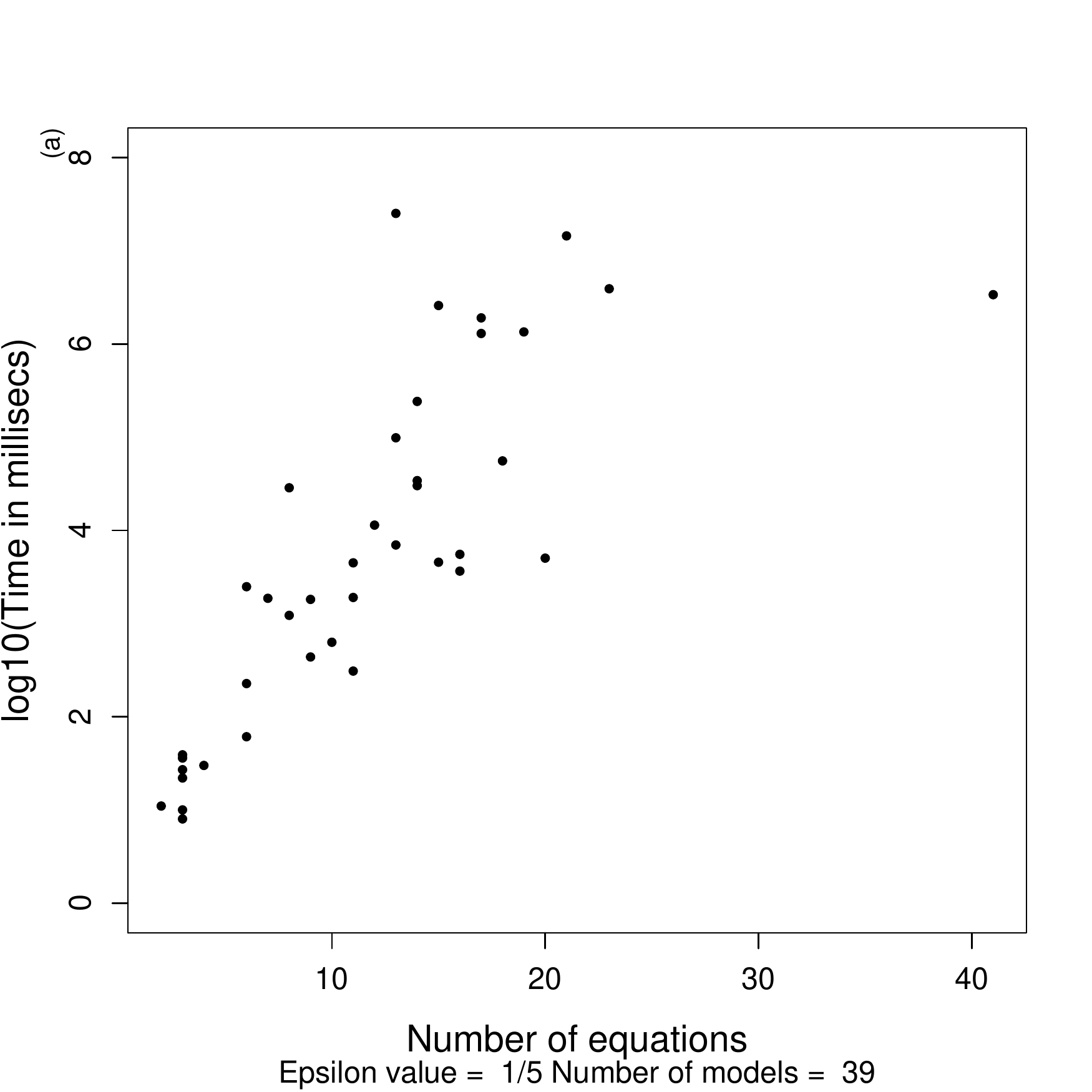}
\includegraphics[width=0.5\textwidth]{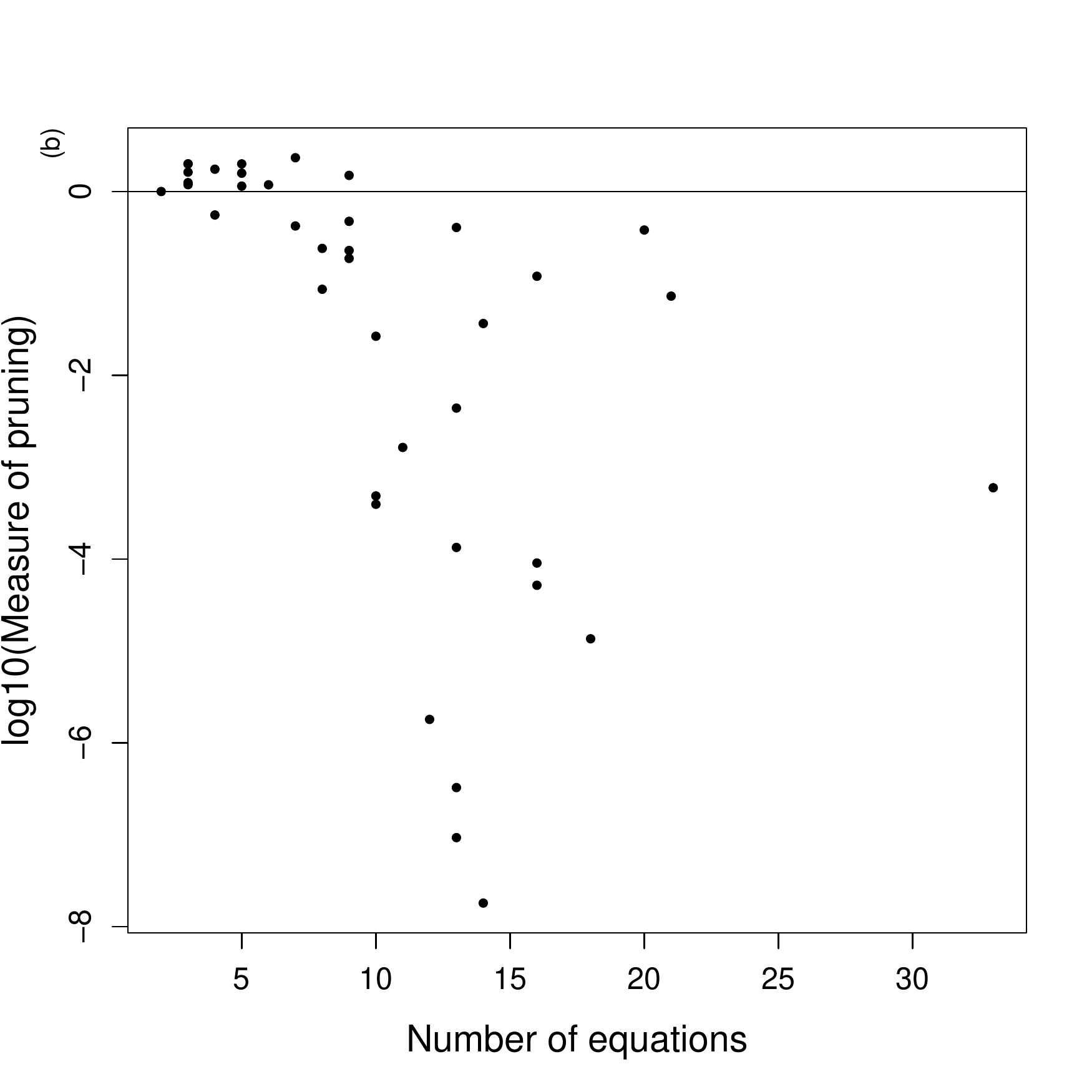}
\caption{(a) Semi-log plot showing log$_{10}$ of CPU running time (in milliseconds) versus
number of equations (which may be greater than number of variables because of conservation laws) for {$\varepsilon =1/5$}. (b) Plot showing the efficacy of tree pruning strategy  for {$\varepsilon = 1/5$}. The scatterplot plots the {logarithm of the measure} of pruning against the number of equations (which may be greater than {the} number of variables because of {the} conservation laws) in Biomodels database. The measure of pruning is computed as per Sect.~\ref{prunningeff}. \label{fig:PrunningBenefits-Timeplot-showing-Time} }
\end{center}
\end{figure}

\begin{figure}
\begin{center}
\includegraphics[width=0.5\textwidth]{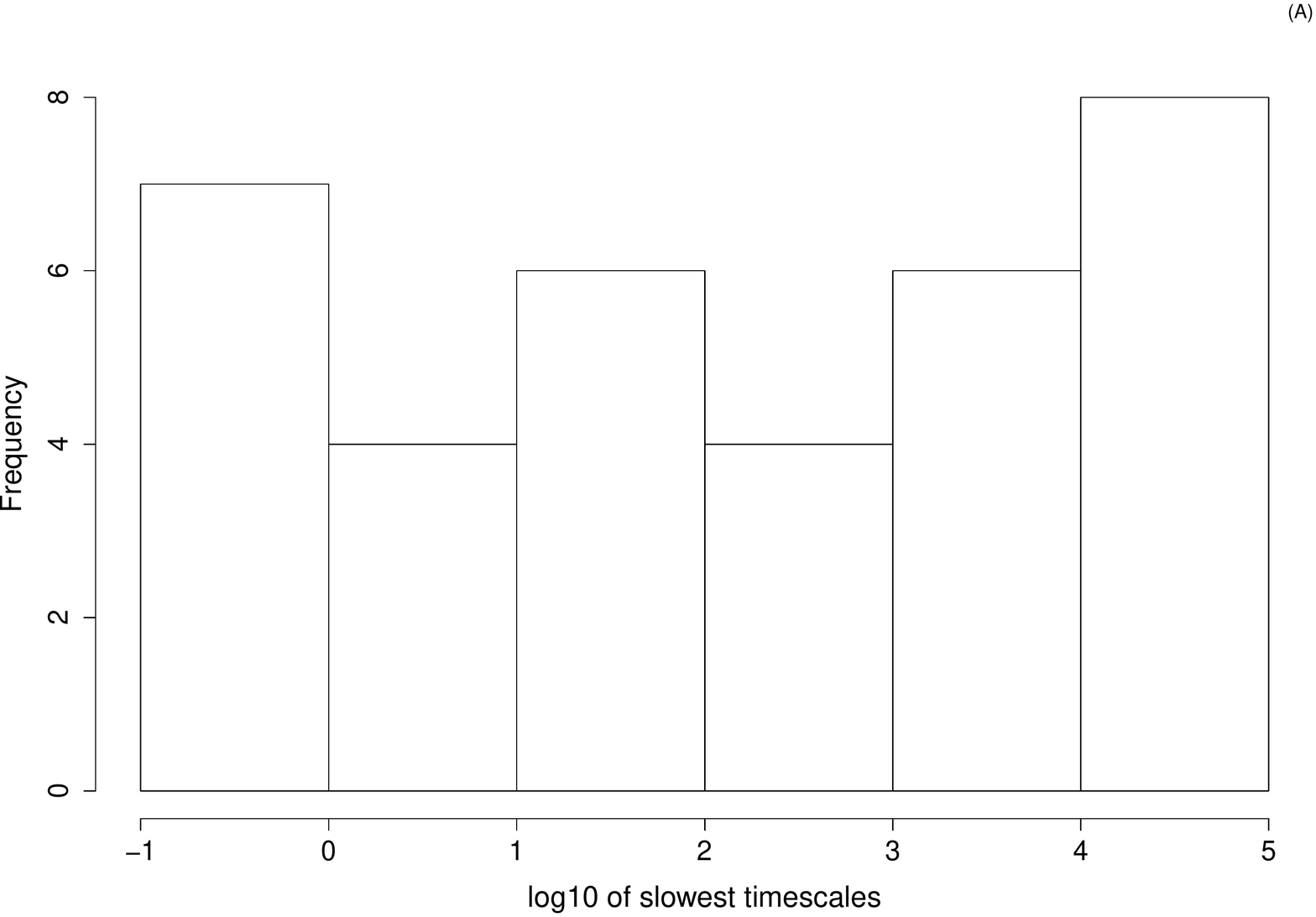}
\includegraphics[width=0.5\textwidth]{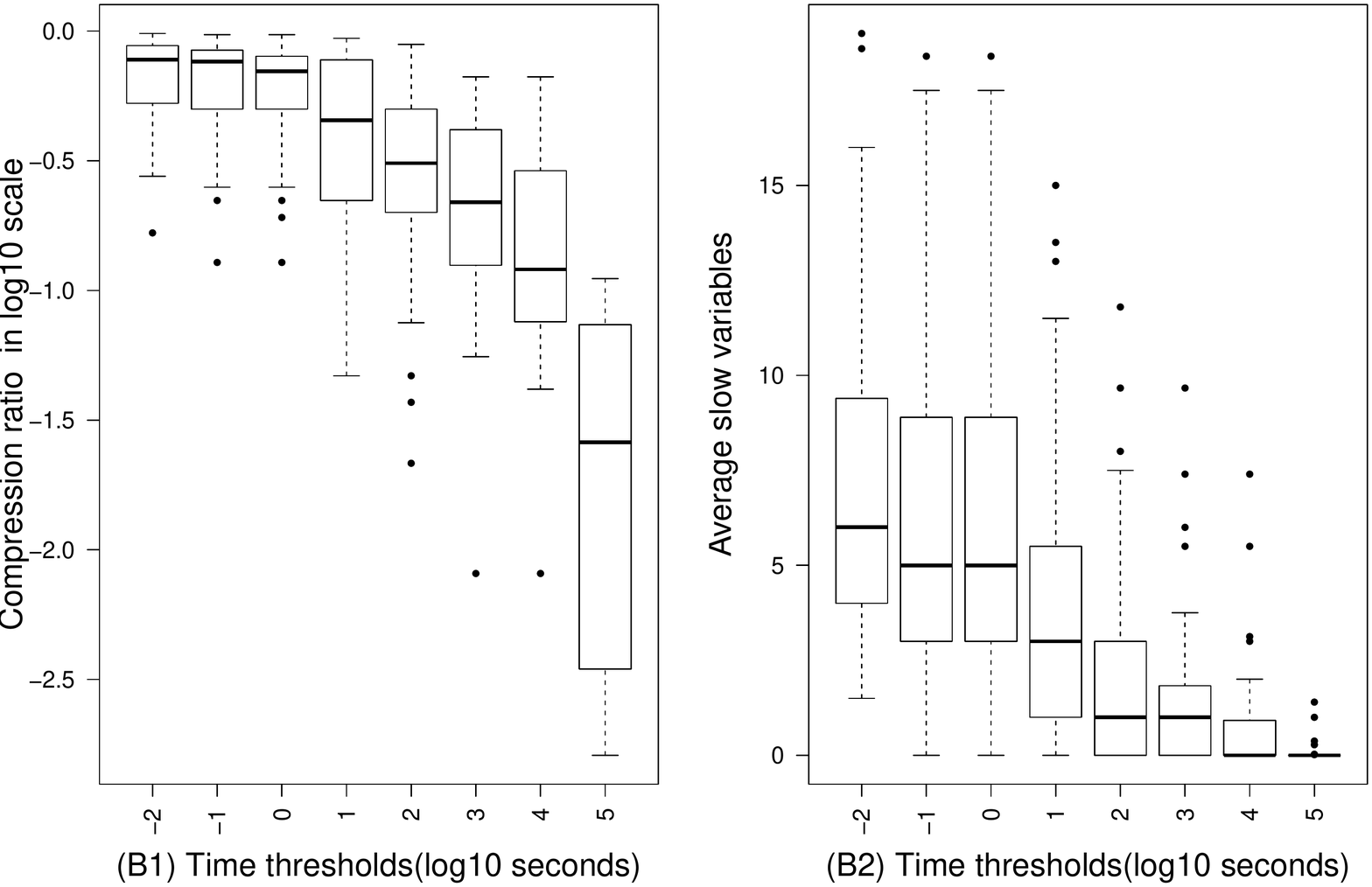}
\caption{\small (A) Histogram showing the distribution of slowest timescales for $35$ models corresponding to $\varepsilon=1/23$. The quasi buffered species (cf. Sect.~\ref{results}) are removed before performing this step.  (B1) and (B2)are boxplots showing compression ratio and average number of slow variables in the Biomodels database
for different values of time threshold $\theta$. The compression ratio is defined as the average number of slow variables $/$ number of variables in the model. The quasi buffered species (cf. Sect.~\ref{results}) are removed before performing this analysis. In (B1), a point represents the compression ratio over all the tropical equilibrations for each model with respect to different time thresholds. Likewise, in (B2), a point represents the average slow variables. The number of slow variables were computed based on rescaled orders ({see} Eq.~\eqref{eq:mut_threshold}) and certain time threshold in seconds. The time thresholds $-2$ to $5$ in the plot are the log$_{10}$ transformed values of time thresholds $0.01,0.1,1,10,100,1000,10000,100000$ in secs. {The boxplot corresponds to  $\varepsilon = 1/23$.} The boxplots of other $\varepsilon$ {values} look similar. \label{fig:Boxplots-showing-proportion-hist_slowtime} }
\end{center}
\end{figure}

 {
Biological models were selected from r25 version of Biomodels \cite{Novere2006}.
For our analysis, we selected models with polynomial kinetics. We
performed two {analyses.} (i) First, we benchmarked our implementation
against models derived from Biomodels. (ii) Second, we computed the average number of slow variables across Biomodels with different time thresholds
to get an estimate of the dimension of {the} invariant manifold.

In the first analysis, the parameters were replaced by the orders
of magnitude according to Eq.~\eqref{scaleparam}. A summary of
the analysis is presented in Table \ref{tab:Summary-of-analysis}.
The analysis is performed to compute all possible combinations of vertices
leading to tropical solutions within a maximal running time of $10,000$ seconds of CPU time. The CPU time threshold for models with $\varepsilon$ value = $1/5$ was further increased to $100,000$ to further ascertain the exponential behaviour.  In practice, we restrict this search space using the tree pruning strategy as explained
in Sect.~\ref{pruning}. While solving the linear inequalities it may happen {that} there exists infinite feasible solutions for a given combination of vertices{. In} such a scenario we report only one solution.
The analysis was repeated with {ten} different values
for $\varepsilon$. A semilog time-plot is presented in Fig.~\ref{fig:PrunningBenefits-Timeplot-showing-Time}
which plots the log of running time in milliseconds versus the number of equations in the
model. It should be pointed out that the number of variables may not be equal to {the} number of equations because of conservation laws which were treated as extra linear equations in our framework.

In the second analysis, the number of slow variables were computed based on the rescaled orders i.e. $\mu_{i} - a_i > \mu_{\text{threshold}}$ and a certain
time threshold as explained in Eq.~\eqref{eq:mut_threshold}. In addition to slow species, we computed the quasi buffered species which are slow variables with very high time threshold. To compute {them} we fixed the timescale threshold to {$100,000$} seconds and the slow species at this threshold are labeled as quasi buffered species. In the model{,} such species are practically constant and in our setting these are subtracted from slow variables. A boxplot is shown in
Fig.~\ref{fig:Boxplots-showing-proportion-hist_slowtime} where a point represents the
compression ratio (i.e. ratio of
average number of slow variables /
total number of variables)
over all the tropical equilibrations for each model with respect to different time thresholds. This was performed for all the models under consideration. We also computed the slowest timescale for each model which is defined as the
smallest
time threshold at which all species become fast (the quasi-buffered species were removed from the model before performing this step). A histogram showing the distribution slowest timescale is presented in Fig.~\ref{fig:Boxplots-showing-proportion-hist_slowtime}. To estimate the slowest timescale, the time threshold is varied and the number of slow species are counted, the threshold at which all species become fast is considered to be the slowest timescale of that model. This histogram indicates that the benchmarked models are representative of a wide variety of cellular processes whose timescales range from fractions of seconds to one day.}

{ The calculations in this section were performed for different
values of the parameter $\varepsilon$. According {to the Eq.\eqref{minplussystem} and} to the geometric interpretation of
tropical equilibrations from Sect.\ref{slowfast} the tropical solutions are either isolated
points or bounded or unbounded polyhedra. Changing the parameter $\varepsilon$
is just a way to approximate the position of these points and polyhedra by
lattices or in other words by integer coefficients vectors.
The approximation results from
the rounding in Eq.\eqref{scaleparam} and better approximations would be to consider
rational (with a largest common denominator), instead of integer orders $\gamma_i$.
Finding the value of $\varepsilon$ that provides
the best approximation is a complicated problem in Diophantine approximation. For that
reason, we preferred an experimental approach consisting in
choosing several values of $\varepsilon$ and checking the robustness of the results. }

\begin{table}
\caption{\small Summary of analysis on Biomodels database\label{tab:Summary-of-analysis}.
 { Tropical solutions here mean existence of at least one feasible solution from all possible combination of vertices of the Newton polytope (in case of infinite solutions, one is picked (cf. Sect. \ref{results})). Timed-out means all solutions could not be computed within $10,000$ seconds of computation time (except for models with $\varepsilon$ value = $1/5$). No tropical solution implies no possible combination of vertices could be found resulting in a feasible solution. {Unit-definition refers to the presence of SBML tag  $<$unitDefinition$>$ which defines the time units of the model to be used in Eq.~\eqref{eq:mut_threshold}. For the models where it is absent seconds is taken to be the default time unit of the model, with the exception of two models whose units have been curated manually by comparison with original papers. }
}
}

\begin{small}
\begin{center}
{
\begin{tabular}{|c|r|r|r|r|r|r|}
\hline
\multicolumn{1}{|p{0.7cm}|}{$\varepsilon$ value} &
\multicolumn{1}{|p{1.2cm}|}{
Total models con\-sidered }&
\multicolumn{1}{|p{1.2cm}|}{{Timed-out models}}&
\multicolumn{1}{|p{1.5cm}|}{{Models without tropical solutions}} &
\multicolumn{1}{|p{1.5cm}|}{{Models with tropical solutions}} &
\multicolumn{1}{|p{1.5cm}|}{ Average running time (in secs)**}&
\multicolumn{1}{|p{1.5cm}|}{ Models with Unit-definition}

\tabularnewline
\hline
\hline
1/5* & 53 & 14 & 0 & 39 & 1354.91 & 12\tabularnewline
\hline
1/7 & 53 & 17 & 0 & 36 & 512.07 & 12\tabularnewline
\hline
1/9 & 53 & 17 & 0 & 36 & 432.54 & 12\tabularnewline
\hline
1/11 & 53 & 16 & 0 & 37 & 756.66 & 12\tabularnewline
\hline
1/19 & 53 & 18 & 2 & 33 & 1063.30 & 12\tabularnewline
\hline
1/23 & 53 & 18 & 1 & 34 & 783.98 & 12\tabularnewline
\hline
1/47 & 53 & 18 & 0 & 35 & 719.79 & 12\tabularnewline
\hline
1/53 & 53 & 19 & 0 & 34 & 387.16 & 12\tabularnewline
\hline
1/59 & 53 & 19 & 1 & 33 & 482.30 & 12\tabularnewline
\hline
1/71 & 53 & 19 & 0 & 34 & 640.32 & 12\tabularnewline
\hline
\end{tabular}
}
\end{center}
*For this $\varepsilon$ the running time threshold was $100,000$ secs.
** For average time computation, the running times of those models which did not timed-out (i.e. $4$th and $5$th column) were considered.
\end{small}
\end{table}

\subsection{Tree Pruning}\label{prunningeff}
In order to evaluate the {efficiency} of tree pruning, we computed the ratio between {the} number of times the linear programming is invoked with {some} tree pruning step  (cf. Fig.~\ref{fig:pruning}) and the possible number of combinations of Newton polytope edges without tree pruning (cf. Eq.~\ref{eq:cart}). This ratio is a measure of efficiency achieved due to pruning. From the computations, it can be seen that for smaller dimensional models the {logarithm} of this ratio is greater than $0$ meaning tree pruning works worse by invoking linear programming more than what is required without tree pruning. However, for {the}  majority of large dimensional models the {logarithm} of this ratio is less than $0$ suggesting significant reduction in the search space due to pruning. The results for $\varepsilon$ value of $1/5$ are presented in Fig.~\ref{fig:PrunningBenefits-Timeplot-showing-Time}.


\subsection{Testing the method}
In order to test the method we consider the cell cycle model
\cite{tyson1991modeling} (referenced as BIOMD00000005 by Biomodels)
in more detail. This is the cell cycle model example analysed
in {Sect.\ref{sub:Example}}. Here we test the detection
of slow and fast species and the accuracy of model reduction.

\subsubsection{Slowness index}
The detection of slow fast species is tested by comparison
with a numerical method introduced in \cite{radulescu2008robust}.
This method consists in simulating trajectories $x_i(t)$ for
each species $i$ of the model and comparing them to the
imposed trajectories $x_i^*$ calculated as solutions
of quasi-steady state equations {cf. Eq.~\eqref{da2}~.} Precisely, $x_i^*$  is the solution of
$\sum_j k_j S_{ij}  \vect{x}^{\vect{\alpha_{j}}} = 0$ in which all species of indices ${l}\neq i$ are replaced by their simulated values  ${x_l}(t)$. Like in
\cite{radulescu2008robust} we use the slowness index $I_i(t) = |\log_{10}(x_i(t)/x_i^*(t))|$
 {(the base of the logarithm is purely conventional)}. Fast species
obey quasi-steady state conditions {(see Eq.~\eqref{da2} and Sect.\ref{slowfast0})}. {Therefore, for fast species,} $I_i$  is close to zero. For slow species, the
trajectories $x_i(t)$ are different from $x_i^*(t)$ and the index $I_i$ is high.
Fig.~\ref{fig:nfkb} shows the values of this index for all the species in the cell cycle
model BIOMD00000005. In our {tropical} method a species is fast or slow depending how the orders $\nu_i=\mu_i - a_i$ compare to a timescale threshold.
For $\varepsilon = 1/9$, we find three tropical solutions, already discussed in Sect.\ref{sub:Example}.
For the solution $\vect{a}_3$
the species 1, 2, and 5 are fast and the species 3, 4, and 6 are slow
(timescales $1 \text{ min}$ or slower).
This solution leads to the reduced model 1 described in Sect.\ref{sub:Example}.
In contrast, species 5 is slow for the two other
equilibrations corresponding to the reduced model 2.
The numerical method based on the slowness index
classifies species 1,2, and 5 as fast and is thus compatible
with the new method for the tropical solution  $\vect{a}_3$ (Fig.~\ref{fig:nfkb}a)).
The reduced model 1 corresponding to the tropical solution $\vect{a}_3$
reproduces with good accuracy the limit cycle oscillations of the cell cycle model
as shown in Fig.~\ref{fig:nfkb}c).

\subsubsection{Accuracy of the reduction}
\begin{figure}
\begin{center}
\includegraphics[width=0.4\textwidth]{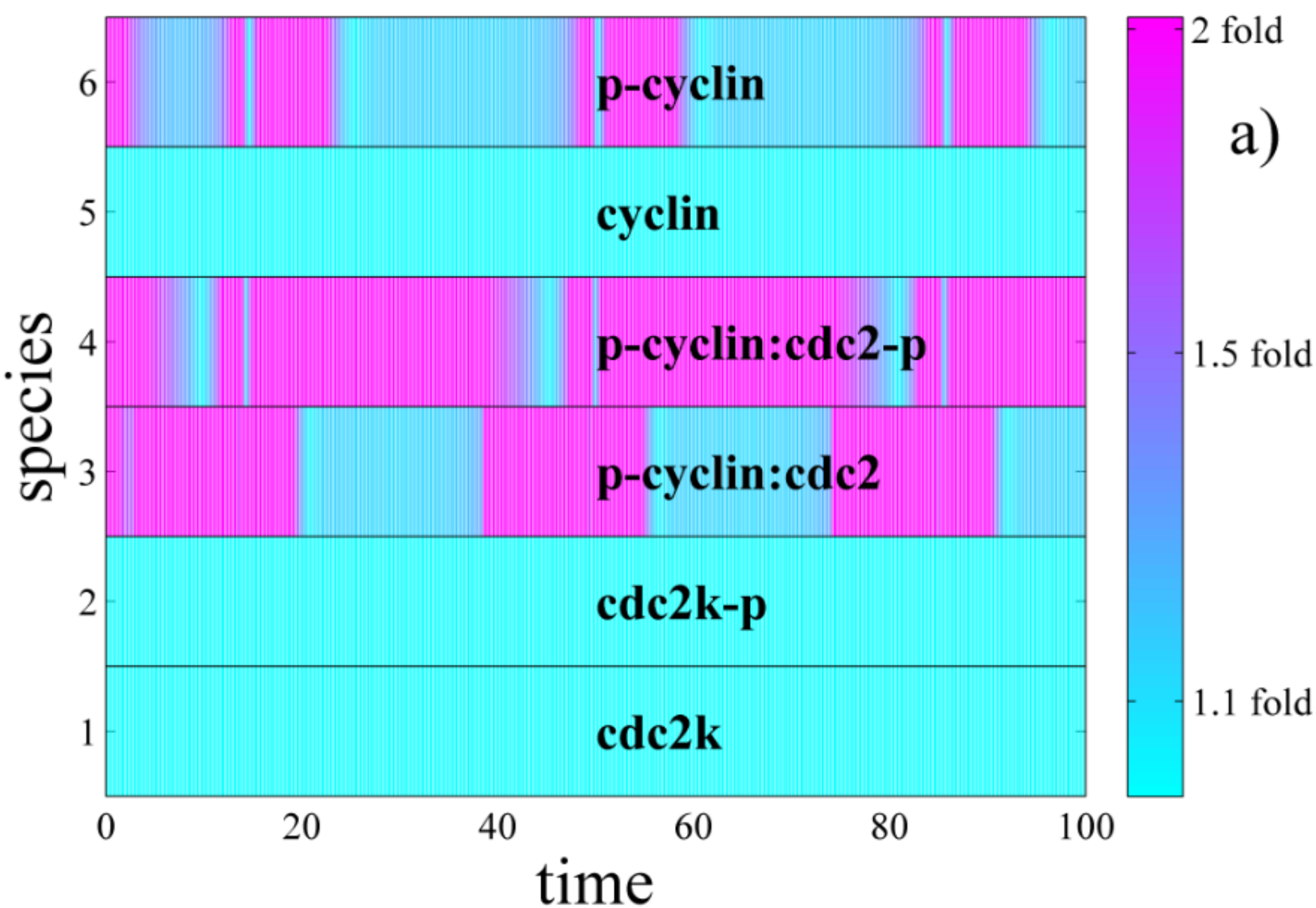}
\includegraphics[width=0.55\textwidth]{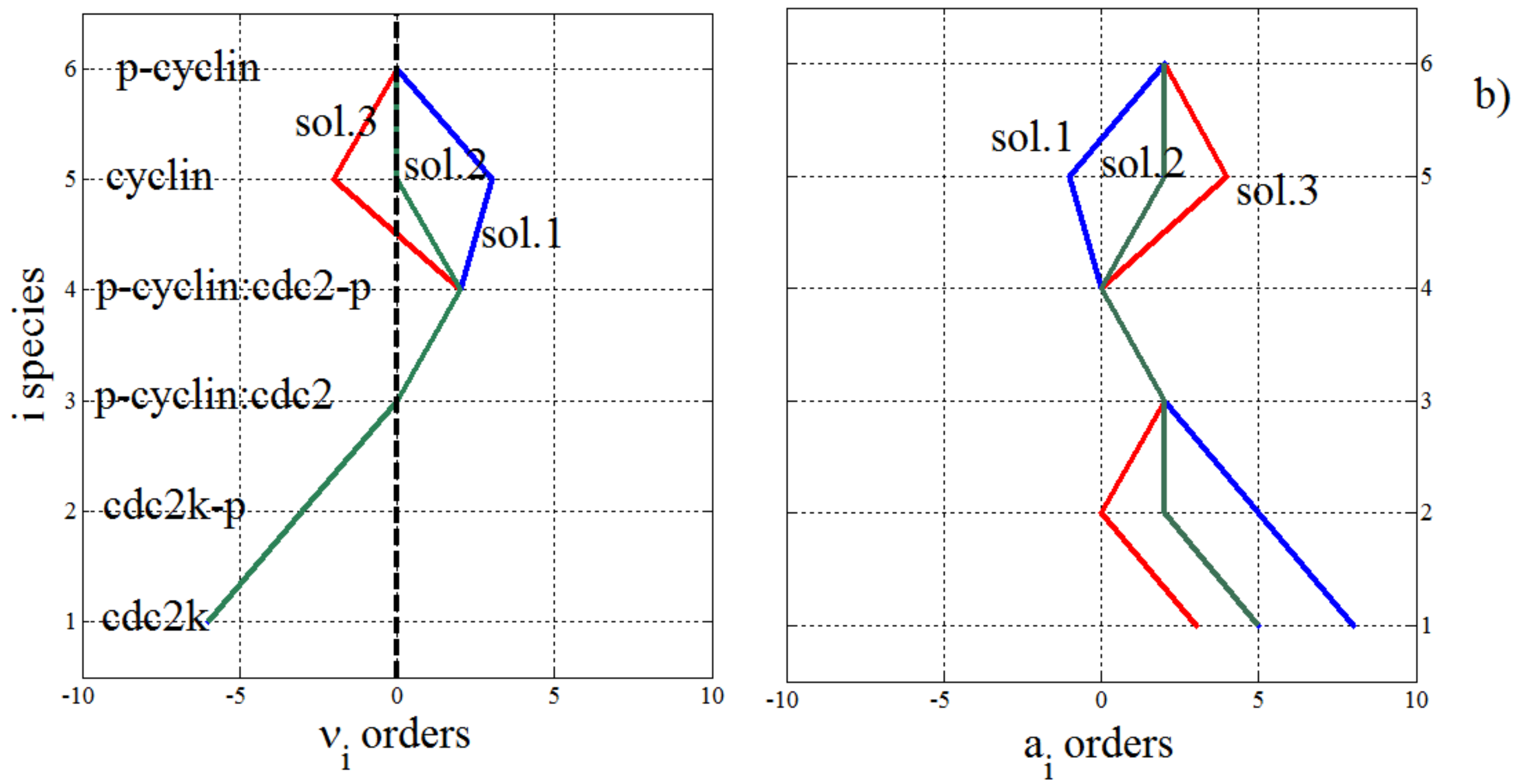}
\includegraphics[width=0.35\textwidth]{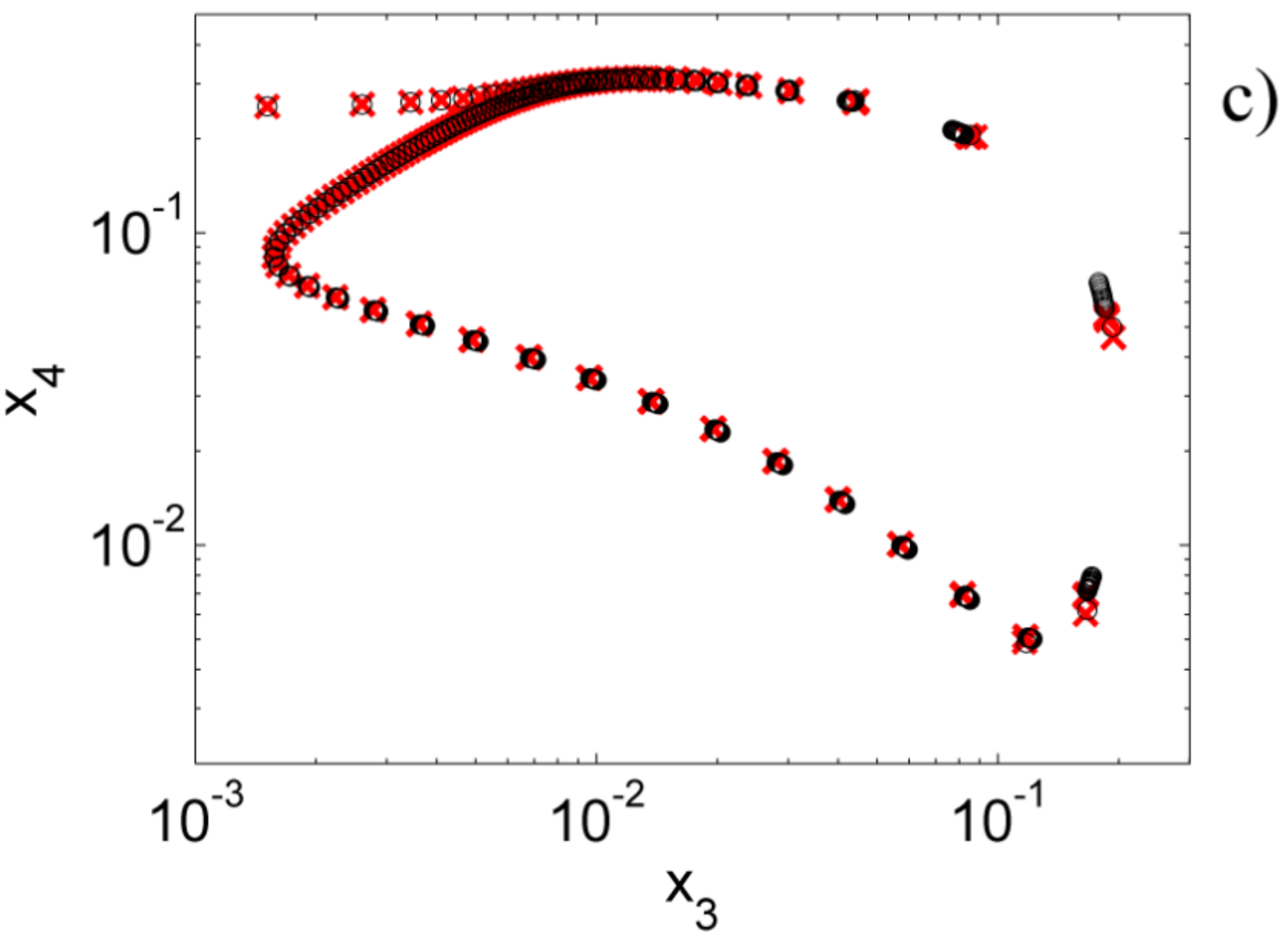}
\includegraphics[width=0.56\textwidth]{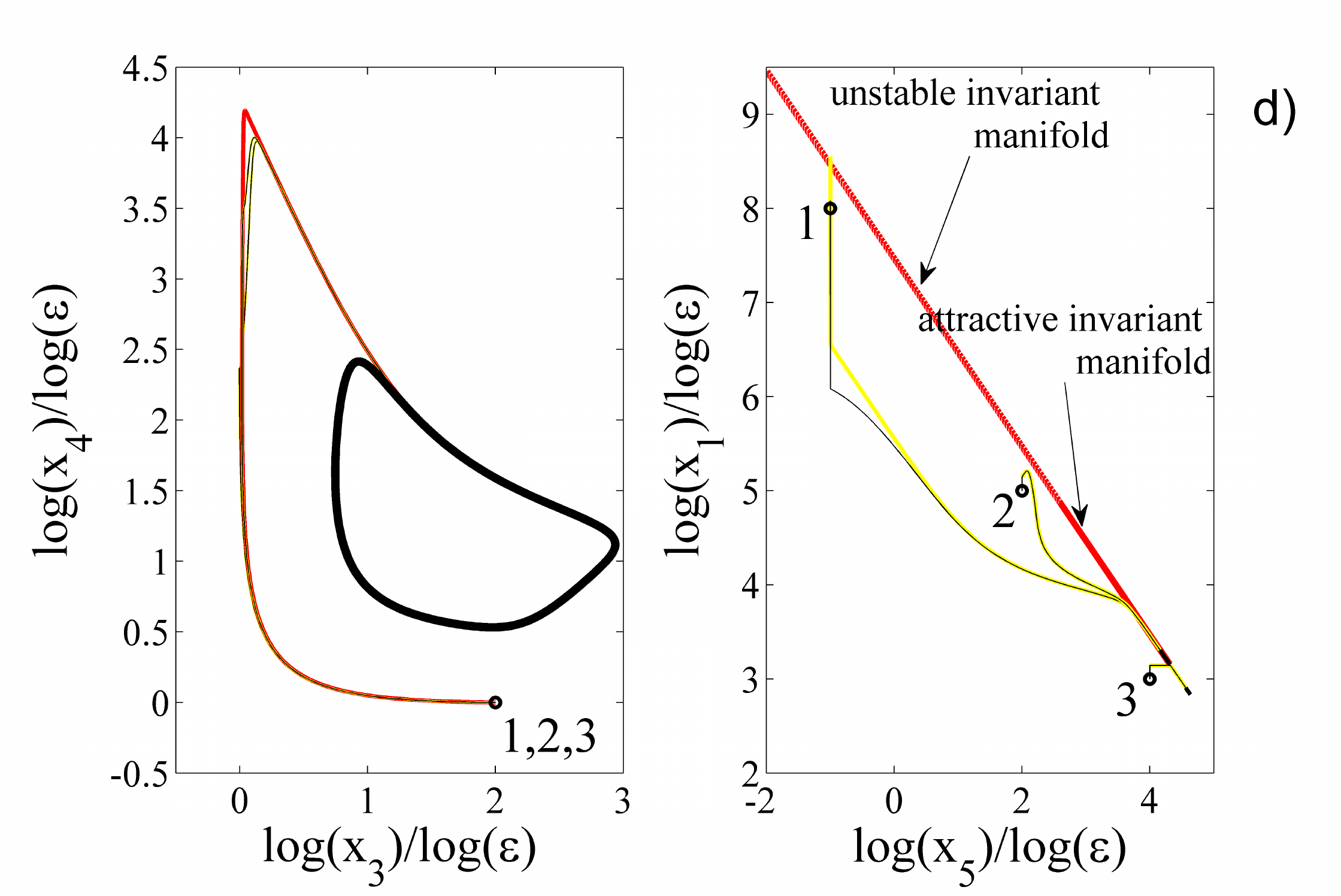}
\end{center}
\caption{\small
Testing tropical slow/fast decomposition and accuracy of reduction of BIOMD00000005 {(cell cycle model)}.
a) The slowness index is represented as a function of time on trajectories:
slow variables have large {slowness index (the fold ratio stands
for the exponentiated
index $exp(|log (x_i(t) / x^*_i (t))|)$)};
b) {Left :} The orders {$\nu_i = \mu_i - a_i$} are represented for different species and for three tropical
solutions. {Cf. eq.\eqref{massactionrescaledtruncated} a
species $i$ evolves on the timescale $\varepsilon^{-\nu_{i}}$ and hence lower $\nu_i$ mean faster
species. The threshold $\theta$ separating slow and fast species set to
$1\,\mathrm{min}$ to satisfy the gap condition \eqref{gapcondition} corresponds to
$\mu_{\text{threshold}} =0$ as defined by \eqref{eq:mut_threshold}. The threshold order
is represented as a dotted line. Fast species have orders below this value, namely
species $x_1$, $x_2$, $x_5$ are fast for the tropical solution $\vect{a}_3$, whereas
only species   $x_1$, $x_2$ are fast for the tropical solutions $\vect{a}_1$ and $\vect{a}_2$.
Right :  Orders $a_i$ for different species and tropical solutions indicate
species concentrations. Cf. eq.\eqref{aorders} higher $a_i$ mean lower concentration. For all
order calculations we have used $\epsilon=1/9$.}
 c) Comparison of the limit cycle trajectories
computed with the full (black circles) and reduced model (red crosses).
d) Model trajectories for the full model (black), reduced model 1 (red) and reduced model 2 (yellow),
starting from three initial data, corresponding to three different tropical equilibrations. The limit
cycle attractor is contained in an invariant manifold. The reduced model 1 provides a good approximation
of the dynamics on the invariant manifold (such as starting from initial data 3), but not outside.
The reduced model 2 is accurate also outside the invariant manifold (see trajectories starting from
equilibrations 1 and 2).
\label{fig:nfkb}
}
\end{figure}

A quantitative estimate of reduction accuracy can be based on the $L^2$ norm
of the difference between trajectories $\vect{x}(t)$, $\vect{x}_{red}(t)$ simulated with the full
and reduced model, respectively.
However, because periods are slightly changed by the reduction, the error could
be defined as $err = inf_{a}\| x(t)-x_{red}(at) \|/\| x(t) \|$, where $a$ is a time scaling
parameter close to $1$. For the trajectories shown in Fig.~\ref{fig:nfkb}c), $err$ is less
than $0.01$ and the optimal scaling parameter is $a = 1.0002$ {(the relative change of the period
is $0.0002$).}

The two other equilibrations lead to {the reduced model 2 that is at least as accurate
as the reduced model 1 (in short, in the reduced model 2, species 5 is considered slow and is not eliminated).}
This reduction accurately {reproduces} the dynamics not only on the limit cycle attractor,
but also when initial data is far from this attractor. This is illustrated in
Fig.~\ref{fig:nfkb}d). We have simulated the full model and the two reduced models
starting from {several initial data $\vect{x}_{0i}, \, i =1,\ldots,3$.}
The initial data of the reduced models is obtained by projection on the corresponding
invariant manifolds. For example, the reduced model 1 evolves on an invariant manifold
whose equations (up to small correcting terms) {are given by \eqref{eq:invman} and read}
$x_5 = k_6/(k_4 x_2)$, $x_1 = k_3 x_2/k_1$.
By computing the eigenvalues of the Jacobian of system \eqref{tyson67}
we found that this invariant manifold has an attractive,
stable {region} (all eigenvalues, except the zero ones corresponding to exact conservation laws,
have negative real parts) and an unstable {region} (where there are eigenvalues with positive real parts).
The  {initial data} vectors $\vect{x}_{01}$ and $\vect{x}_{02}$
are close to the unstable {region} of the invariant manifold. Therefore,
trajectories starting from these initial data {first get away from the manifold and after} large
excursions {approach} the attractive part of the manifold.
Reduced model 2 is able to reproduce these transients but not the
reduced model 1 (Fig.~\ref{fig:nfkb}d) because the latter is valid only on the
{slowest}
attractive
invariant manifold.

\subsection{{Comparison with COPASI time separation} method}

We compared our proposed method against the existing
{tool COPASI \cite{Hoops15122006,Surovtsova01112009}.
{COPASI is a software for simulation and analysis of biochemical networks. This software accepts
and generates several model exchange formats including the widely spread systems biology markup language (SBML)
format and is very popular in the computational biology community.
To the best of our knowledge, COPASI is the only major biochemical networks tool that implements
time separation of variables. To accomplish this aim
COPASI proposes a modified ILDM (intrinsic low dimensional  manifold) method.} This
method computes slow and fast modes which are transformations of species concentrations as described in \cite{deuflhard1996dynamic,zobeley2005new,Surovtsova01112009}. More precisely,
COPASI performs a Schur block decomposition of the Jacobian matrix $\vect{J}$, consisting in finding a non-singular matrix
$T$ such that $T^{-1} J T = \begin{pmatrix}  S_{slow} & 0 \\ 0 & S_{fast} \end{pmatrix}$, where $S_{slow}, S_{fast}$ have real Schur form, i.e. they are upper triangular matrices with possibly non-vanishing
elements  on the first subdiagonal.
The time threshold (needed to separate the slow and fast blocks of the Jacobian matrix)  is automatically captured in this method by finding a gap in the spectrum of the Jacobian (cf. Sect.\ref{slowfast0})}.

\begin{figure}
\begin{center}
\includegraphics[width=0.90\textwidth]{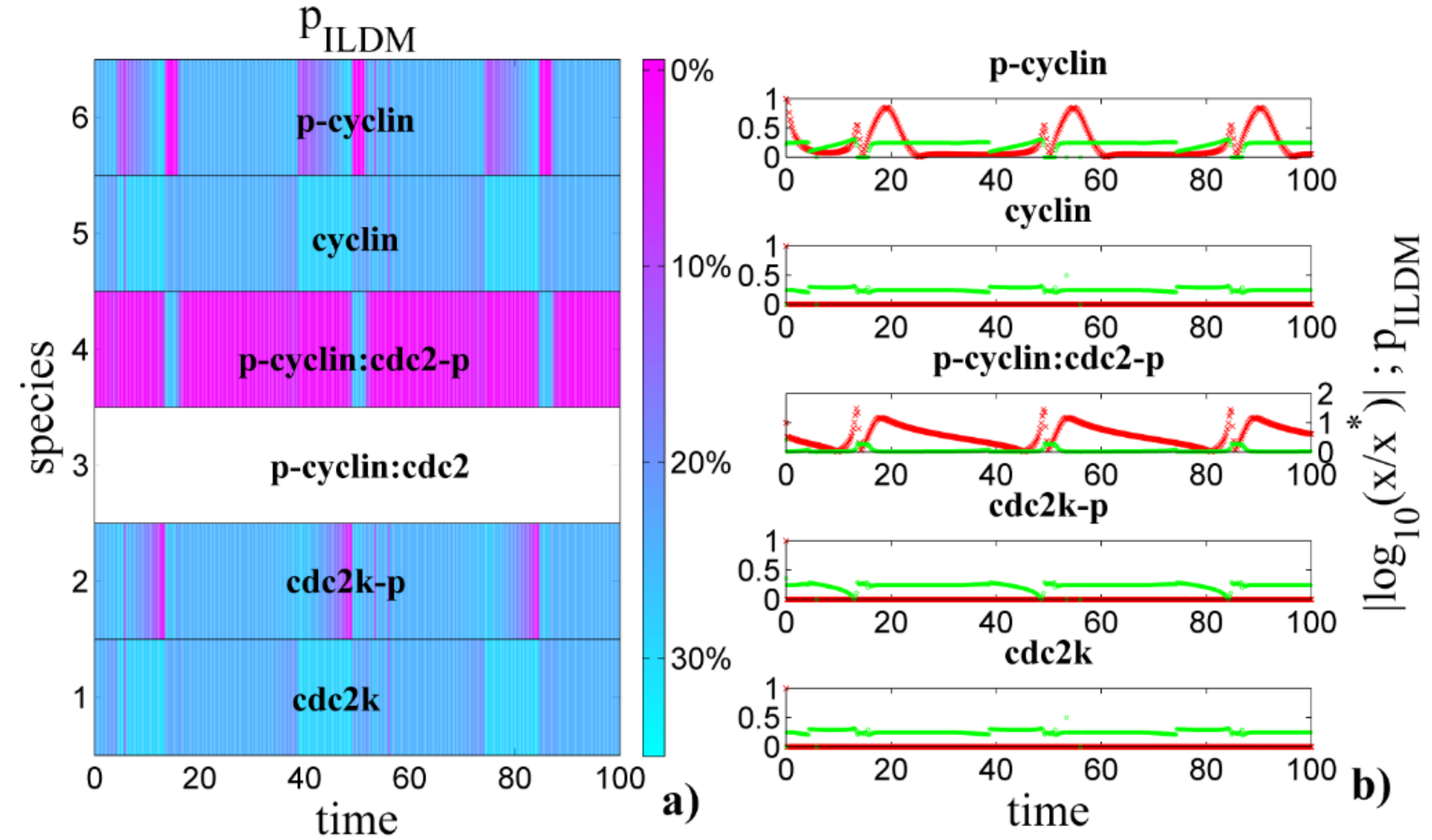}
\caption{\small Summary of the analysis of model BIOMD00000005 using ILDM method and comparison with our method based on the slowness index.
\label{fig:ILDM}
{
The model was simulated in COPASI
{from $0$ to  $100$ min with}
 default initial concentrations. The species $3$ is eliminated using the single conservation law.
For all the remaining species we represent the time dependence of their contributions
$p_i$, where $0 \leq p_i \leq 1, \, \sum_i p_i = 1$ to the fast space. The fractions $p_i$ are generated by COPASI and the values $p_i$ are color coded
in the left panel a).
Chemical species with largest contribution to the fast space (largest $p_i$) are supposed to be fast species (as explained in \cite{Surovtsova01112009}). Therefore $p_i$ (in green in right panel b)) and our slowness index $I_i= |\log_{10}(x_i/x_i^*)|$ (in red) should be anti-correlated.
This is well verified for species $1,4,5$ ($p_i$ are relatively high when $I_i$ are close to zero, and close to zero when $I_i$ are relatively high, but it is not well verified for species $6$ (p-cyclin) {whose slowness index has large peaks in places where the COPASI contribution to fast modes stays constantly high}.
}
}
\end{center}
\end{figure}

In order to compare the modified ILDM method against {our tropical and slowness index methods,} we computed the fast space of the model using COPASI for $100$ time steps between $1$ and $100$ min and checked the contribution of each species to this fast space.
{ COPASI defines the contribution of a species $i$ to a mode $j$ as the matrix element $T^{-1}_{ji}$.}
These contributions of various species to fast modes are provided by COPASI as fractions $p_i$,
where $i$ is {the species index}, $0 \leq p_i \leq 1, \, \sum_i p_i = 1$.
COPASI declare species with largest contribution to the fast space
(largest $p_i$) as fast species. For exactly the same trajectory, we have computed the values of the
slowness indices ${I}_i = | \log_{10} ( x_i(t) / x^*_i (t) ) |$. For fast species, {$I_i$} should be close to zero.
{Fig.~\ref{fig:nfkb}a) and b) summarizes the comparison between the slowness index and the tropical
method. The tropical solution $\vect{a}_3$ leads to the reduced model 1 (cf. Sect.\ref{sub:Example}) and copes with the limit cycle
trajectory used in this test. The timescale orders $\nu_i$ of the variables for this tropical solutions
identify species $x_1$, $x_2$, $x_5$ as fast and species $x_3$, $x_4$, $x_6$
as slow (see also Sect.\ref{sub:Example}). As can been seen in Fig.~\ref{fig:nfkb}a) the slowness index of species
$x_1$, $x_2$, $x_5$
is close to zero
for all times. The slowness index of species $x_3$, $x_4$, $x_6$
can reach large values. Therefore the tropical method and the slowness index method provide exactly the
same timescale decomposition.
COPASI time separation can not be compared directly to the tropical method, because it generates a timescale decomposition that changes with time and which is valid for a trajectory. However, it can be compared with the slowness index decomposition.
Fig.~\ref{fig:ILDM} summarizes the comparison between COPASI and the slowness index. It should be noted that the species $x_3$ is automatically eliminated by COPASI using the single conservation law present in the model.
The slowness index and COPASI contribution to fast space should be anticorrelated: when the first one is
small the latter should be big and vice versa. Species $x_1$ and $x_5$ have high contribution towards the fast
space and very low slowness index (see Fig.~\ref{fig:ILDM}b).
For these species we can say there is consistence between COPASI and slowness index.
Species $x_2$ also has large contribution to fast space except for some
intervals where COPASI may classify it as slow. Our method unambiguously classifies this species
as fast (cf. Fig.~\ref{fig:ILDM}b) its slowness index is very low for all times).
Most importantly,
COPASI fails to identify correctly time intervals where species $6$ is slow as indicated
by the large value of the slowness index co-existing with large values of contribution $p_i$ (as large as for species $1$ and $2$, see Fig.~\ref{fig:ILDM}b). According to COPASI this species is similar
to {the} fast species $x_2$, whereas our methods indicate it {is} similar to the slow species $x_4$ and $x_3$. }
{As a matter of fact, COPASI determines slow variables by comparing the values of contributions $p_i$
to the fast and slow modes.
 {Despite the existence of} a spectral gap, the differences of the indices $p_i$ between species that are considered slow and fast can be relatively small and therefore
  {this}
  classification is not robust. In contrast, our methods
classify species in a robust way. Indeed, we directly associate timescales to species, via the orders
$\nu_i$ and these timescales are well separated for slow and fast species.
 Using these orders, we found that the fastest slow species $x_3$ and $x_6$ are $100$ times slower than the slower fast species $x_5$.
Furthermore, as shown in Fig.\ref{fig:nfkb}a, our slowness index is very sensitive to differences in timescales. Fast species  $x_1$, $x_2$, $x_5$ keep this
index low for all times, whereas the corresponding COPASI indices are not always high. Generally, it should not be recommended to use species contributions to modes as indicative of their timescales, as COPASI does.
For instance, the sum of two or more species can be a slow mode, even if all these species are fast (cf. Sect.
\ref{sub:reduction}, this situation is typical for fast cycles). The fast species have in this case high contributions to a slow mode which may qualify them as slow according to the
{species}
contribution criterion.
}

\section{Conclusion}
{We have addressed the problems of timescale decomposition and
model reduction of biochemical networks. Our approach relies  on the notion
of tropical equilibration. Tropical equilibrations represent a
generalization of steady states and
correspond to compensation of dominant fluxes acting on species
concentrations. The remaining, uncompensated weaker fluxes are responsible for the
slow dynamics of the system on attractive invariant manifolds. The
correspondence between tropical equilibrations and attractive
invariant manifolds has been exploited here to associate, to each tropical
equilibration, a reduced model. We show elsewhere \cite{SamalGrigorievRadulescu2015a} that
when there is an infinity of tropical equilibration solutions, these
can be organised into branches, each branch corresponding to the same
reduced model.

We have proposed an algorithm  to compute tropical equilibrations.
We have used this algorithm to determine the fast/slow partition of
chemical species in a network of biochemical reactions that is the first
and often critical step in model reduction algorithms.
In particular, the number of slow species provides the size
of minimal dynamic models
to which complex biochemical networks can be reduced. The validity of our
reductions depends on concentration and parameter orders, as well as on initial
data.
For the simple example of Michaelis-Menten kinetics we obtained validity
conditions for various reductions as inequalities among orders of magnitude
of concentrations and parameters. These validity conditions define
large domains in the concentrations and parameters spaces.
Previous work on larger models suggests a larger applicability of this result
which implies that the resulting reductions are robust \cite{radulescu2008robust}.

The benchmarking of our algorithm
on the Biomodels database shows that a significant
dimension compression can be performed on cell dynamics models
at timescales of $1000 s$ and larger. Starting
with complex models having more than 30 variables, minimal models have
median numbers of 2 slow variables. This suggests that, at least
piece-wise in parameter and phase space, the tasks fulfilled
by molecular networks are relatively simple.
The need for having complex machineries with many regulators
to perform simple tasks (such as relaxation to
steady states or limit cycle oscillations) could be justified
by system robustness. A system having a very large number of
variables and parameters, multiple timescales
and only a few slow degrees of freedom
is generically robust with respect to perturbations of variables and
parameters \cite{gorban2007dynamical}.

Our methods can be also
used to study sensitivity issues and identifiability of parameters from trajectories.
A parameter is sensitive if changing its value induces large changes of model trajectories.
In our analysis of Tyson's model we have seen that some parameters
of the full model are not present in the reduced model. Although
the orders of magnitude of these parameters are important (changing
them may change the reduction), small changes of their values
do not lead to changes of the model trajectories.
Parameters of the full model that are not present in the reduced
model are therefore insensitive. It may also happen that parameters
of the reduced model are combinations (for instance multivariate monomials)
of the parameters of the full model. If these combinations are sensitive,
then so are the parameters they contain. However, parameters that are parts of such
combinations can not be determined independently
from the observed trajectories, which leads to parameter
non-identifiability  issues \cite{radulescu2012frontiers}. Thus,
insensitive parameters and
parameter lumping resulting from model reduction
can be used to asses local identifiability of
system parameters  \cite{radulescu2012frontiers}.
The idea of relating parameter lumping and parameter identifiability
can also be found in other computational algebraic geometry approaches
\cite{meshkat2009algorithm}.

Solving the tropical equilibration problem and finding a slow-fast decomposition is the first step
for model reduction. The remaining steps consist in elimination of the fast variables by solving
systems of algebraic equations. We have shown how this can be performed for  simple examples.
In the case of more complex models, the elimination can be performed numerically, or symbolically.
Tropical methods can simplify this task by replacing the full  systems by tropically truncated systems.
In particular, the binomial or toric case when the truncation has only two monomials is particularly interesting because
for this case there are rapid algorithms for computing steady states \cite{millan2012chemical}.
Higher approximation can be provided by Newton-Puiseux expansions \cite{radulescu2015},
that encompass tropical solutions in their lowest order.
Although the calculations needed for formal reduction could be long, once the model is reduced, it can be used in various applications, such as a part of larger networks, or in models of tissues
and organisms where the same biochemical network has to be replicated in several interacting cells.
Furthermore, our reductions have a strong geometrical basis. In future work, we will exploit this property to show how to endow the reduced model with a reaction network structure and how to identify inclusion relations among different reduced models.

Several other open questions will be addressed in future work. For instance, our current
algorithm finds the tropical equilibrations for fixed values of the parameters.
It would be very interesting to formally classify all the possible
reductions and phase portraits of a reaction network with a given topology and reaction
rates, for all possible values of the parameters. We have solved this problem by hand for the Michaelis-Menten kinetics.
In the future we will extend our algorithms in order to compute how the tropical equilibration
solutions depend on  parameters.
For this purpose we will extend techniques used for {\em linear
quantifier elimination}
\cite{Weispfenning:88,DolzmannSturm:97a,WeberSturm2011a}
and incorporate them into Algorithm~\ref{algo:1}.
}

\subsubsection*{Acknowledgements.}
This work has been supported by the French ModRedBio CNRS Peps, and EPIGENMED
Excellence Laboratory projects. D.G. is grateful to the
Max-Planck Institut f\"ur Mathematik, Bonn for its hospitality
during writing this paper and to  Labex CEMPI (ANR-11-LABX-0007-01).

%
%


\end{document}